\documentclass[fleqn,usenatbib]{rasti}

\usepackage{newtxtext,newtxmath}

\usepackage[T1]{fontenc}

\usepackage{wasysym}
\usepackage{chemformula}

\DeclareRobustCommand{\VAN}[3]{#2}
\let\VANthebibliography\thebibliography
\def\thebibliography{\DeclareRobustCommand{\VAN}[3]{##3}\VANthebibliography}

\usepackage{graphicx}	
\usepackage{amsmath}	

\definecolor{colblue}{rgb}{0.012, 0.262, 0.875}

\title[WALTzER]{WALTzER: Wide-band Atmospheric Laboratory for Transiting Exoplanet Research}

\author[A. G. Sreejith et al.]{
A. G. Sreejith$^{1}$\thanks{E-mail: sreejith.aickara@oeaw.ac.at },
L. Fossati$^{1}$,
M. Lendl$^{2}$, 
P. E. Cubillos$^{1,3}$, 
C. Broeg$^{4,5}$, 
D. Ehrenreich$^{2}$,
B.-O. Demory$^{5,6,4}$,
\newauthor 
D. Piazza$^{4}$,
I. Argyriou$^{7}$,
V. Bourrier$^{2}$,
M. Braam$^{5}$,
L. Carone$^{1}$,
Y. Carteret$^{2}$,
H. Chakraborty$^{8}$,
A. Deline$^{2}$,
\newauthor 
O. Demangeon${^9}$,
C. Farret Jentink$^{2}$,
G. Giono$^{1}$,
Ch Helling$^{1}$,
T. Jeřábková$^{10}$,
P. Kabáth$^{11}$,
T. A. van Kempen$^{12}$,
\newauthor 
A.~Su\'{a}rez~Mascare\~{n}o$^{13,14}$,
J. C. Morales$^{15,16}$,
M. Munari$^{17}$,
C. Opitom$^{18}$,
R. Ottensamer$^{19}$,
G. Parma$^{2}$,
L. Roth$^{20}$, 
\newauthor 
J. Skottfelt$^{21}$,
R.-S. Taubner$^{1}$,
A. Tanzer$^{1}$,
T. G. Wilson$^{8}$,
B. Edwards$^{12}$,
G. Scandariato$^{22}$,
V. Van Grootel$^{23}$,
\newauthor 
J. Venturini$^{2}$,
A. Vorburger$^{3}$,
S. Ambily$^{24}$,
A. Brandeker$^{25}$,
T. Belenguer$^{26}$,
E. Bolmont$^{2,27}$,
S. Boro Saikia$^{19}$,
\newauthor 
H. M. Cegla$^{7,28}$,
Q. Changeat$^{29}$,
C. Danielski$^{30,17}$,
K. Fahrion$^{19}$,
B. Flemming$^{31}$,
K. France$^{31}$,
E. Gillen$^{32}$,
\newauthor 
K. Hakim$^{33,34}$,
C. Haswell$^{35}$,
R. Helled$^{36}$,
M. Ivchenko$^{20}$,
J. Jacobs$^{23}$, 
K. Kislyakova$^{19}$,
V. Korol$^{12}$,
T. Kupfer$^{37}$,
\newauthor 
A. Kuzucan$^{2,27}$,
J. McCormac$^{8}$,
C. Mordasini$^{4,5}$,
J. Munday$^{38}$,
G. Nelemans$^{39}$,
A. F. Pala$^{40}$,
I. Ribas$^{15,16}$,
\newauthor 
J. Sanz-Forcada$^{41}$,
C. Snodgrass$^{18}$,
S. Sousa$^{9}$,
J. Žák$^{11}$,
Y. Alibert$^{4,5}$,
S.P.D. Borthakur${^1}$,
S. R. Carberry Mogan$^{20}$,
\newauthor 
J.~M.~Carrasco$^{42,43,16}$,
C. Dorn$^{44}$,
J. P. D. Gow $^{21}$,
M. Güdel$^{19}$,
F. Kerschbaum$^{19}$,
D. Kitzmann$^{5,6,4}$,
\newauthor 
G. Laky$^{1}$,
E. McDonald$^{19}$,
H. Ottacher$^{1}$,
P. A. Sossi$^{45}$,
V. Trigler$^{1}$,
R. Wallner$^{1}$
\\
$^{1}$Space Research Institute, Austrian Academy of Sciences, Schmiedlstrasse 6, 8042 Graz, Austria\\
$^{2}$Département d’Astronomie, Université de Genève, 51 Chemin Pegasi, 1290 Versoix, Switzerland\\
$^{3}$INAF -- Osservatorio Astrofisico di Torino,
    Via Osservatorio 20, 10025 Pino Torinese, Italy\\
$^{4}$Space Research and Planetary Sciences, Physics Institute, University of Bern, Sidlerstrasse 5, 3012 Bern, Switzerland\\
$^{5}$Center for Space and Habitability, University of Bern, Gesellschaftsstrasse 6, 3012 Bern, Switzerland\\
$^{6}$ARTORG Center for Biomedical Engineering Research, University of Bern, Bern, Switzerland\\
$^{7}$Institute of Astronomy, KU Leuven, Celestijnenlaan 200D, box 2401, 3001 Leuven, Belgium\\
$^{8}$Department of Physics, University of Warwick, Gibbet Hill Road, Coventry CV4 7AL, UK\\
$^{9}$Instituto de Astrofísica e Ciências do Espaço, Universidade do Porto, CAUP, Rua das Estrelas, 4150-762 Porto, Portugal\\
$^{10}$Department of Theoretical Physics and Astrophysics, Faculty of Science, Masaryk University, Kotlářská 2, 611 37 Brno, Czechia\\
$^{11}$Astronomical Institute, Czech Academy of Sciences, Fričova 298, 251 65 Ondřejov, Czech Republic\\
$^{12}$SRON Netherlands Institute for Space Research, Niels Bohrweg 4, 2333 CA Leiden, The Netherlands\\
$^{13}$Instituto de Astrofísica de Canarias (IAC), 38205 La Laguna, Tenerife, Spain\\
$^{14}$Departamento de Astrofísica, Universidad de La Laguna (ULL), 38206 La Laguna, Tenerife, Spain\\
$^{15}$Institut de Ciències de l’Espai (ICE, CSIC), c/ de Can Magrans s/n, Campus UAB, 08193 Bellaterra, Spain\\
$^{16}$Institut d’Estudis Espacials de Catalunya (IEEC), 08860 Castelldefels, Barcelona, Spain\\
$^{17}$INAF--Osservatorio Astrofisico di Arcetri, Largo E. Fermi 5, 50125 Firenze, Italy\\
$^{18}$Institute for Astronomy, University of Edinburgh, Royal Observatory, Edinburgh EH9 3HJ, UK\\
$^{19}$Department of Astrophysics, University of Vienna, Türkenschanzstrasse 17, A-1180 Vienna, Austria\\
$^{20}$KTH Royal Institute of Technology, Stockholm, Sweden\\
$^{21}$Centre for Electronic Imaging, School of Physical Sciences, The Open University, Milton Keynes MK7 6AA, UK\\
$^{22}$INAF--Osservatorio Astrofisico di Catania, Via S. Sofia 78, 95123 Catania, Italy\\
$^{23}$Space Sciences, Technologies and Astrophysics Research (STAR) Institute, Université de Liège, 19C Allée du 6 Août, 4000 Liège, Belgium\\
$^{24}$Silicon Austria Labs GmbH, Sandgasse 34, A-8010 Graz, Austria\\
$^{25}$Department of Astronomy, Stockholm University, AlbaNova University Center, 10691 Stockholm, Sweden\\
$^{26}$Departamento de Óptica Espacial, Instituto Nacional de Técnica Aeroespacial (INTA), Carretera de Ajalvir km 4, 28850 Madrid, Spain\\
$^{27}$Centre pour la Vie dans l’Univers, Université de Genève, Genève, Switzerland\\
$^{28}$Centre for Exoplanets and Habitability, University of Warwick, Gibbet Hill Road, Coventry CV4 7AL, UK\\
$^{29}$Kapteyn Astronomical Institute, University of Groningen, 9747 AD Groningen, The Netherlands\\
$^{30}$Departament d’Astronomia i Astrofísica, Universitat de València, Av. Vicent Andrés Estellés 19, 46100 Burjassot, Spain\\
$^{31}$Laboratory for Atmospheric and Space Physics, University of Colorado Boulder, UCB 600, Boulder, CO 80309, USA\\
$^{32}$Astronomy Unit, Queen Mary University of London, Mile End Road, London E1 4NS, UK\\
$^{33}$Institute of Astronomy, KU Leuven, Leuven, Belgium\\
$^{34}$Royal Observatory of Belgium, Brussels, Belgium\\
$^{35}$School of Physical Sciences, The Open University, Walton Hall, Milton Keynes MK7 6AA, UK\\
$^{36}$Department of Astrophysics, University of Zurich, Winterthurerstrasse 190, 8057 Zurich, Switzerland\\
$^{37}$Hamburger Sternwarte, University of Hamburg, Gojenbergsweg 112, 21029 Hamburg, Germany\\
$^{38}$Institute for Physics and Astronomy, University of Potsdam, Karl-Liebknecht-Str. 24/25, 14476 Potsdam, Germany\\
$^{39}$Department of Astrophysics/IMAPP, Radboud University, P.O. Box 9010, 6500 GL Nijmegen, The Netherlands\\
$^{40}$European Southern Observatory, Karl-Schwarzschild-Str. 2, 85748 Garching bei München, Germany\\
$^{41}$Centro de Astrobiología, CSIC-INTA, Madrid, Spain\\
$^{42}$Departament de Física Quàntica i Astrofísica (FQA), Universitat de Barcelona (UB), Martí i Franquès 1, E-08028 Barcelona, Spain\\
$^{43}$Institut de Ciències del Cosmos (ICCUB), Universitat de Barcelona (UB), Martí i Franquès 1, E-08028 Barcelona, Spain\\
$^{44}$Institute for Particle Physics and Astrophysics, ETH Zurich, CH-8093 Zurich, Switzerland\\
$^{45}$Department of Earth and Planetary Sciences, ETH Zurich, CH-8092 Zurich, Switzerland\\
}
\date{Accepted XXX. Received YYY; in original form ZZZ}

\pubyear{2026}

\begin{document}
\label{firstpage}
\pagerange{\pageref{firstpage}--\pageref{lastpage}}
\maketitle

\newpage
\pagebreak
\begin{abstract}
Understanding the origin and evolution of planetary atmospheres, and the connection between their lower and upper atmospheric layers, remains a fundamental challenge in exoplanet science. Addressing these questions requires observations spanning a broad wavelength range, from the ultraviolet to the near-infrared, together with the ability to simultaneously characterise planetary and stellar variability. The Wide-band Atmospheric Laboratory for Transiting Exoplanet Research (WALTzER) is an ESA F3 mission candidate designed to address this need through simultaneous low-resolution spectroscopy and high-precision photometry of transiting exoplanets. WALTzER covers simultaneously the near-ultraviolet (NUV: 240--325 nm) and visible (VIS: 450--820 nm) bands with two medium resolution spectrographs and the near-infrared (NIR: 900--1700 nm) range with a broadband photometer. We outline the main science drivers and present the mission concept and instrument design. WALTzER's broad, simultaneous wavelength coverage enables the detection and characterisation of the atmospheres of more than 30 gas-rich exoplanets and provides data relevant for a broad range of Solar System and Galactic astrophysics science cases. The payload consists of a 35 cm Cassegrain telescope feeding three co-aligned channels (i.e. NUV, VIS, NIR), sharing a common optical train to minimise instrumental systematics, while maximising observational efficiency. 
\end{abstract}




\section{Introduction}

The discovery of exoplanets has opened an exciting new field in astrophysics and has completely transformed our understanding of planets. Transiting exoplanets, that is planets passing in front of their host star as seen from Earth, enable comprehensive studies of the atmosphere of these foreign worlds through transmission spectroscopy. More than 6000 exoplanets are currently known and thousands more are expected to be soon found by space-based surveys \citep[e.g. TESS, PLATO;][]{ricker2015,rauer2025} explicitly dedicated to the search for transiting planets orbiting bright stars, and thus providing opportunities for detailed atmospheric characterisation.

Gas-rich transiting exoplanets are key laboratories for investigating planet formation and evolution. Owing to their close proximity to their host stars, they also provide extreme environments in which to study atmospheric composition and fundamental atmospheric processes such as cloud formation and atmospheric escape. Infrared (IR) and visible observations probe the lower atmosphere ($\sim$1--10$^{-6}$ bar), enabling the reconstruction of its chemical inventory \citep[e.g.][]{kempton2024} and the characterisation of ubiquitous aerosols \citep[e.g.][]{pont2013,kreidberg2014,Sing2016}. Transmission spectra of some planets show clear atmospheres with prominent Na ($\sim$580 nm), K ($\sim$770 nm), and H$_2$O ($\sim$1.4 $\mu$m) absorption features, while others surprisingly lack one or both of the Na and K signatures \citep[e.g.][]{Sing2016}. The hottest planets can additionally exhibit strong TiO features, which are thought to contribute to the formation of atmospheric temperature inversions \citep[e.g.][]{Sedaghati2017,chen2021_TiO}. Ultraviolet (UV) observations, in contrast, probe the extended upper atmosphere ($<$10$^{-6}$ bar), where atoms and ions are continuously lost to space \citep[e.g.][]{vidalmadjar2003,Fossati+2010,haswell2012,ehrenreich2015,Sing2019}. Atmospheric escape is expected to play a key role in the long-term evolution of planetary atmospheres and may therefore be an important driver of the observed diversity of exoplanetary systems \citep[e.g.][]{owen2016,kubyshkina2022,Guillot2023,snellen2025_review}.

Despite the complementary information provided by observations at different wavelengths, a coherent characterisation of exoplanet atmospheres remains challenging because existing observations are inevitably incomplete in wavelength coverage and/or disjointed in time. Even in the few cases where both the lower and upper atmospheres have been observed, the measurements have generally been obtained at different epochs and with different facilities, and are therefore subject to different instrumental systematics and potentially different stellar activity states. Combining such heterogeneous observations can complicate the construction of a consistent picture of the atmospheric structure and evolution, and can introduce additional uncertainties into their interpretation \citep[e.g.][]{giacobbe2021,xue2024,blain2024,bachmann2025,verma2025,finnerty2025}. For example, the \textit{James Webb Space Telescope} (JWST) is enabling studies of atmospheric composition, thermal structure, and molecular signatures in the near- and mid-IR, while ESA's \textit{Ariel} mission \citep{Tinetti2018} will provide a large statistical survey of transiting exoplanet atmospheres. In parallel, the \textit{Pandora} SmallSat \citep{Hoffman2022} aims at disentangling stellar variability from planetary atmospheric signals through simultaneous visible (imaging) and near-IR (spectroscopy) observations. However, their capabilities are somewhat limited. The general observatory nature of \textit{JWST} limits the number of targets that can be observed and its instruments do not cover the blue-optical and UV bands, while \textit{Ariel} and \textit{Pandora} lack optical and UV spectroscopic coverage. This calls for a stable and efficient facility simultaneously covering the UV to the near-IR, further enabling connecting its data with those of other facilities covering longer wavelengths. WALTzER is designed to probe the lower (near-IR and visible) and upper (near-UV) atmospheres of transiting exoplanets with a single instrument, providing a temporally and instrumentally coherent view of their atmospheric properties. 

WALTzER will simultaneously obtain near-UV (NUV; 240--325 nm) and visible (450--820 nm) transmission spectra at a resolving power of R$\sim$3000, plus near-IR (NIR; 900--1700 nm) high-precision photometry of more than thirty transiting gas-rich exoplanets. The space-based platform and mission architecture dedicated to the characterisation of exoplanetary atmospheres ensures that WALTzER will survey a significant number of transits ($\approx$20) of each system, thus uniquely enabling the investigation of variability and its origin. WALTzER will increase the number of systems with UV-to-IR observations by a least a factor of ten, while uniquely providing simultaneous UV-to-IR wavelength coverage. 
Its broad wavelength coverage and spectral resolution will give simultaneous access to tracers probing lower and upper atmospheric structure and composition (e.g. Fe{\sc i}, Fe{\sc ii}, Mg{\sc i}, and Mg{\sc ii} in the NUV; Na{\sc i}, K{\sc i}, and TiO in the optical) and enables constraining the altitude and inferring the composition of aerosols, while the near-IR (NIR) imager will cover a number of water bands. As part of its core survey, WALTzER will provide the observational material needed to answer the mission's core science questions:
\begin{enumerate}
    \item What is the origin of planetary atmospheres and how do they evolve?
    \item How do lower and upper atmospheric layers interact?
\end{enumerate}

Next to its core science case, WALTzER's characteristics make it a versatile instrument for exploring a wide range of science cases spanning from the Solar System to Galactic astrophysics. Therefore, as part of consortium science, WALTzER will also monitor the radiation environment of nearby planet-hosts, investigating implications for habitability, and study active bodies in the Solar System, specifically long-period comets, interstellar objects, and the moons of giant planets. In addition, 25\% of the time of the nominal mission will be available to the community.

This paper is organised as follows. Section~\ref{sec:science} presents the mission's science case split into core survey (Section~\ref{sec:core_survey}); additional consortium surveys (Section~\ref{sec:additional_consortium_surveys}); and possible community science (Section~\ref{sec:community_science}). Section~\ref{sec:requirements} gives the instrument and mission top-level requirements needed to reach the science goals. WALTzER's instrument and science operations are presented in Section~\ref{sec:instrument} and \ref{sec:science_operations}, respectively. Section~\ref{sec:conclusions} provides a summary of the mission concept and gathers the conclusions.

\section{WALTzER's science}\label{sec:science}

\subsection{Core survey}\label{sec:core_survey}


Key processes contributing to the observed exoplanet diversity include 
\begin{itemize}
    \item accretion of gas or planetesimals,
    \item atmospheric escape/loss driven by absorption of the stellar X-ray and UV radiation (together XUV),
    \item and orbital migration through interactions within the disk and/or with the host star and outer companions
\end{itemize} 
\citep[e.g.][]{mordasini2024}. These processes leave observable traces in the exoplanets’ atmospheric composition, vertical structure, and mass loss \citep[e.g.][]{oberg2011,venot2016,lothringer2021,kubyshkina2022,molliere2022}. Gas-rich exoplanets are the only exoplanets whose atmospheres can be robustly characterised and consequently are landmark objects to map planet formation and evolution, allowing detailed insights into the processes shaping them.


Ground- and space-based observations revealed the complexity of atmospheric physics at low altitudes, notably eastward winds and temperature inversions caused by shortwave absorbers \citep[e.g. TiO, VO; e.g.][]{Sing2016,kempton2024,snellen2025_review}. Furthermore, aerosols are common at low altitudes and impact the atmospheric chemistry and energy budget \citep[e.g.][]{helling2019_review}. Upper atmospheres ($<$10$^{-6}$\, bar), observed primarily in the UV, often show strong mass loss detectable through H, He, and metal lines \citep[e.g. Fe, Mg; e.g.][]{Fossati+2010,Sing2019}. This process is driven by stellar XUV radiation absorbed in the thermosphere, which can shape planetary evolution by altering atmospheric structure, for example through the preferential escape of lighter species \citep[differentiation; e.g.][]{lammer2003,lammer2025}. The exoplanet population clearly shows the impact of atmospheric evolution. For example, few objects exist in the ``Neptunian desert'', which is a region of the parameter space where planets cannot retain their atmospheres \citep[e.g.][]{mazeh2016}. Indeed, objects near this desert show strong escape from their extended atmospheres \citep[e.g.][]{ehrenreich2015,bourrier2018}.

WALTzER is designed to study upper and lower atmospheres jointly. This is important because lower atmosphere properties (winds, thermal profile, chemistry, aerosols) can profoundly affect the upper atmospheric temperature, chemistry, and escape \citep[e.g.][]{koskinen2022,steinrueck2023}. For example, aerosols directly impact the boundary conditions for upper-atmospheric escape models at $\sim$1 $\mu$bar \citep[e.g.][]{lavvas2019,huang2023}. Aerosols may also sequester from the gas phase metals \citep[e.g. Mg and Fe; e.g.][]{2019A&A...626A.133H,2021A&A...648A..80H,2023A&A...671A.122H} that can contribute to the energy budget \citep[e.g.][]{fossati2021,fossati2023,Fossati2025}. Consequently, to fully understand the connection between lower and upper atmosphere we must collect complete atmospheric profiles from simultaneous observations probing all accessible altitudes. WALTzER's mid-resolution ($\approx$3000) NUV and VIS spectrographs probe the upper and lower atmosphere, respectively, while the NIR photometer probes the lower atmosphere, further linking WALTzER’s observations to those of JWST and Ariel \citep{Tinetti2018}.

\begin{figure}
	\includegraphics[width=\columnwidth]{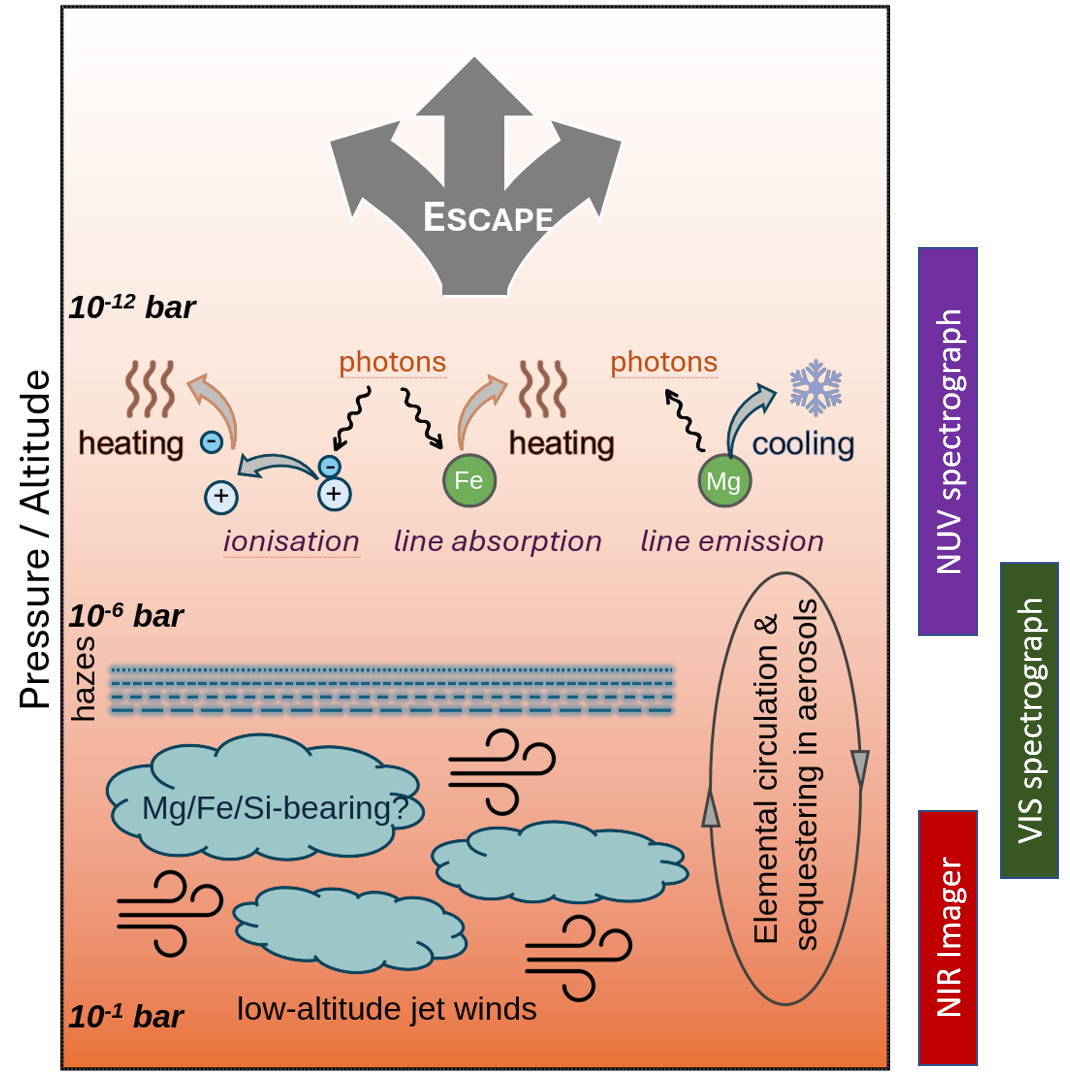}
    \caption{Illustration of the main features and processes probed by WALTzER in the atmosphere of a gas-rich exoplanet. The vertical bars on the right hand side indicate the approximate location of the atmosphere probed by each instrument channel.}
    \label{fig:illustration_atmosphere}
\end{figure}

\subsubsection{Planetary origins, evolution, and the role of atmospheric escape}
\label{sec:escape} 

Planets emerge from the protoplanetary disc with extended hydrogen or steam envelopes that are highly susceptible to mass loss \citep[e.g.][]{owen2020_review,thao2024,barat2025}. At the same time, stellar activity is strongest at early ages \citep[e.g.][]{johnstone2021} leading to high XUV irradiation and frequent exposure to high-energy events such as flares and coronal mass ejections \citep[e.g.][]{ilin2025}. These stellar phenomena episodically enhance mass loss and alter atmospheric chemistry \citep[e.g.][]{venot2016,hazra2022,feinstein2022}. After an initial phase of strong escape \citep[e.g.][]{owen2016,fossati2017,kubyshkina2018}, mass loss can continue to gradually erode the planetary atmosphere and may alter its composition throughout the planet’s entire life \citep[e.g.][]{kubyshkina2022,lammer2025}. The strength of the escape is controlled by the temperature in the upper atmosphere, which is set primarily by stellar emission and activity, but also by the heat flux from the lower atmosphere and by the abundance of metals acting as heating (e.g. Fe; via line absorption) and cooling (e.g. Mg; via line emission) agents \citep[e.g.][]{lammer2003,owen2016,Fossati2025,misener2026}. WALTzER has been designed to simultaneously observe, and thus constrain, these properties.


\subsubsection{Planetary climates and the role of clouds}
\label{sec:clouds} 

Being tidally locked, hot gas-rich planets possess continuously irradiated day sides and likely cloudy night sides \citep[e.g.][]{helling2019_review}. The lower atmosphere is dominated by a global wind jet \citep{showman2002} that shifts the planet’s hottest spot to the east of the substellar point \citep{knutson2007}. The wind flow in the lower atmosphere apparently gives rise to vertical flows that feed escape through the upper atmosphere, connecting the two domains \citep{seidel2019}.

Numerous transmission spectra revealed the presence of aerosols \citep[e.g.][]{Sing2016}, which dampen atomic and molecular features. Aerosol occurrence is not uniform: while some planets show thick clouds \citep[e.g.][]{alam2020}, others appear to be cloud-free \citep[e.g.][]{nikolov2018} or show spectral slopes corresponding to scattering from small haze particles \citep[e.g.][]{pont2013}. Cloud formation in hot atmospheric environments is poorly understood \citep[e.g.][]{sindel2022}. This lack of understanding is problematic, as clouds strongly affect the atmosphere’s chemical and thermal properties. For example, cloud formation can remove the heating/cooling agents Fe/Mg from the gas phase in the upper atmosphere, a process that may shed light on how cloud formation initiates. At the same time, clouds alter the planet’s albedo controlling how much, and where, stellar radiation is absorbed \citep[e.g.][]{helling2019_review,Lee2023,carone2026,Bisht2026}. This leads to changes in the upper atmospheric temperature and, for some planets, in the mass loss rate, and thus in the overall atmospheric evolution \citep[e.g.][]{lavvas2021,koskinen2022}.

\subsubsection{The role of WALTzER in exoplanetology}
\label{sec:role}

WALTzER is designed to simultaneously probe the lower and upper atmosphere of close-in gas giants (Figure~\ref{fig:illustration_atmosphere}). It will
\begin{itemize}
    \item constrain the lower atmosphere’s temperature and detect the presence of metal oxides, alkali metals, and aerosols;
    \item detect Mg and Fe, determining from their feature amplitudes whether the atmosphere is actively escaping by measuring Roche lobe overflow;
    \item and measure stellar activity (through e.g. Mg{\sc ii} h\&k, H$\alpha$).
\end{itemize}
Consequently, WALTzER will link escape from the upper atmosphere to conditions in the lower atmosphere and the stellar environment for its prime sample of 54 planets, sampling different ages, masses, and orbital periods (Section~\ref{sec:targets}).

WALTzER will do even more for the youngest planets, aged $\leq$50 Myr. Stars at this age are so active that the state-of-the-art technique to measure planetary masses (radial velocity) fails to deliver accurate measurements \cite[e.g.][]{cale2021,zicher2022,suarez_mascareno2022,sikora2023}. 
JWST transmission observations of young planets have been able to measure precise planetary masses from their atmospheric scales heights \citep[e.g.,][]{ BaratEtal2025ajV1298TaubTransitNIRSpec, DaiEtal2026arxivV1298TaueNIRSpecCS2, MurphyEtal2026ajV1298cTransitJWST}. For the young system V1298 Tau, masses from transmission spectroscopy and TTV analyses \citep[][]{LivingstonEtal2026naturV1298TTVs} seem to agree with each other better than with RV estimates \citep{SuarezMascarenoEtal2022natasV1298TauRadialVelocities, FinocietyEtal2023mnrasV1298TauRV, DamassoEtal2026arxivV1298TauRVs}.
WALTzER transmission spectra will constrain those masses from the shape and amplitude of the spectral features \citep{dewit2013,dewit2025}. This will be key to the interpretation of these objects, for example by constraining the planetary bulk density (see Section~\ref{sec:complementarity}).

WALTzER will also homogeneously study cloud formation and its impact on upper atmospheres by 
\begin{itemize}
    \item constraining the altitude of the main cloud deck via the attenuation of the Na and K lines in the visible band,
    \item deriving the density and particle sizes of the cloud precursors via scattering slopes apparent across the NUV-visible-NIR bands,
    \item and determining whether the presence and altitude of aerosols correlates with the strength of the escape from the upper atmosphere.
\end{itemize}

For the brightest handful of targets, WALTzER will also collect full-orbit phase curve observations measuring the planetary brightness from different angles, which provides complementary information to the transmission spectra. The NIR channel measures the planetary thermal emission, giving temperature at different longitudes, which constrains atmospheric circulation regimes \citep{knutson2007}. Thermal emission in the VIS channel can be corrected for using the NIR data, while the visible and NUV channels constrain reflected starlight \citep[e.g.][]{hooton2022}. From the phase curve shape, the location of clouds in the atmosphere can be inferred, while the spectral properties constrain aerosol composition \citep[e.g.][]{sudarsky2000,evans2013,mayorga2019}.

\subsubsection{Target list}
\label{sec:targets}

\begin{figure*}
	\includegraphics[width=\textwidth]{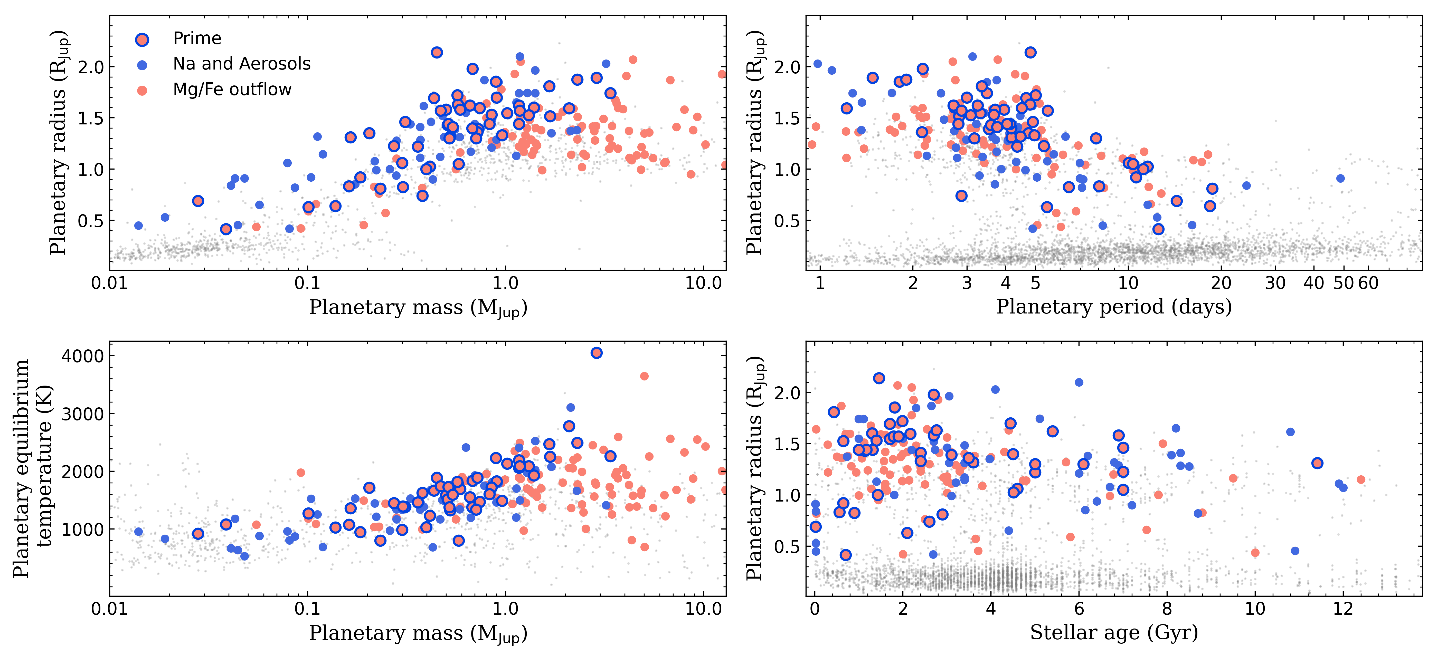}
    \caption{WALTzER target sample (coloured) against the full exoplanet population (grey). WALTzER targets corresponding to gas-rich exoplanets with radii above $\sim$0.4 R$_{\rm Jup}$. Blue refers to targets for which Na and aerosols can be detected, while pink refers to targets for which escape can be detected (see Section~\ref{sec:requirements}).}
    \label{fig:targets}
\end{figure*}

WALTzER's core survey will range from young, low-density, Neptune-mass objects, particularly susceptible to atmospheric loss, to ultra-hot Jupiters (UHJs), which are outstanding targets for detailed atmospheric characterisation. More exoplanets will be known by launch (e.g. from PLATO) and thus the final target list will be set before launch. A fully suitable target sample, however, is already available (Figure~\ref{fig:targets}).

\begin{figure*}
	\includegraphics[width=\textwidth]{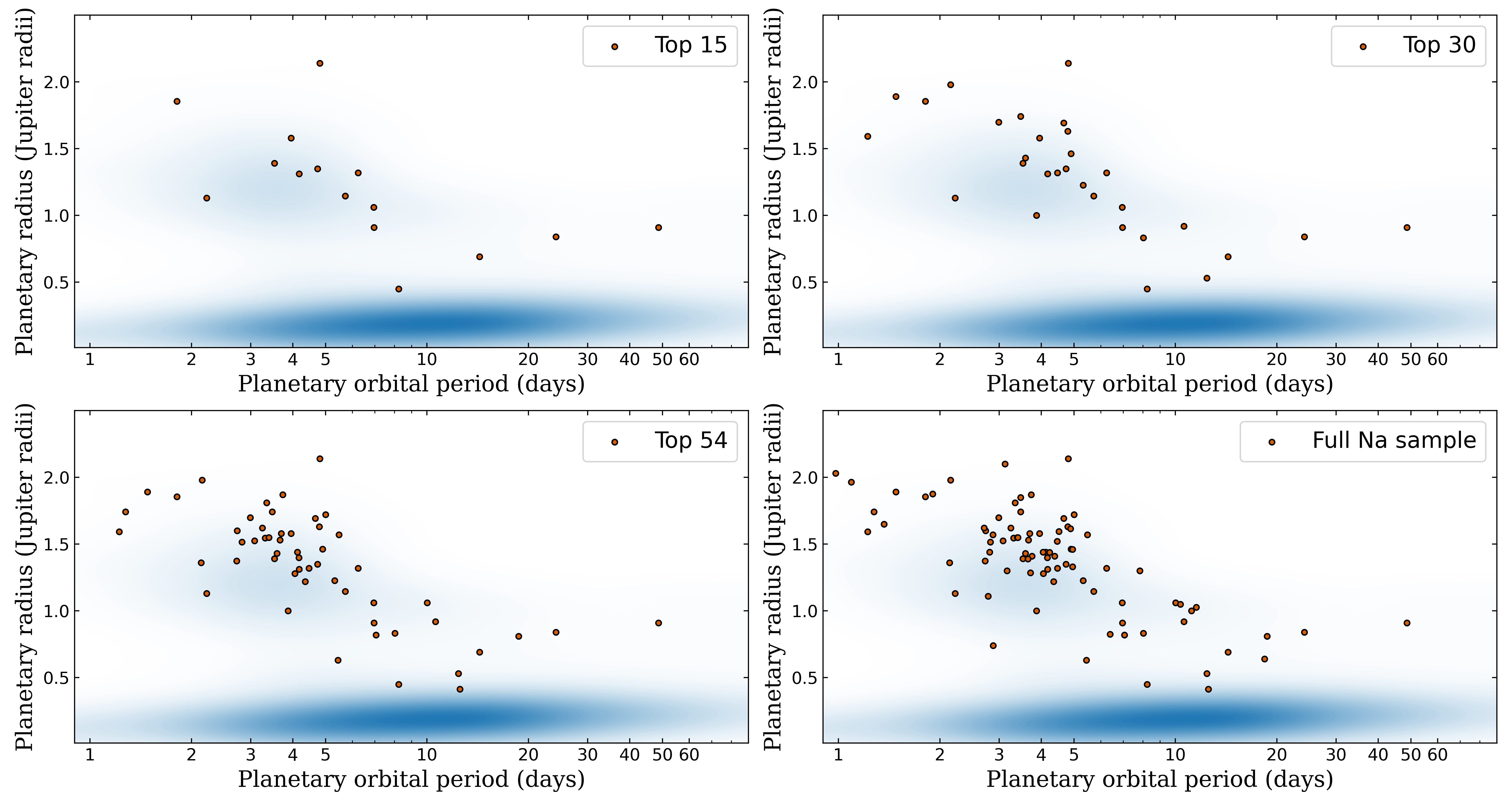}
    \caption{Top, from left to right: position of the best 15 and 30 WALTzER targets (circles) in the planetary radius vs orbital period plane compared to the population of known exoplanets indicated by the blue density map. At least 30 targets are required to uniformly cover the parameter space. Bottom, from left to right: same as top, but for the prime sample (54 targets) and for the Na detection sample (84 targets).}
    \label{fig:targets_differentN}
\end{figure*}

As shown in Figure~\ref{fig:targets_differentN}, at least 30 targets are required to sample the parameter space covered by the known transiting gas-rich planets. Cross matching the science requirements, the instrument characteristics, and the known sample of transiting gas-rich planets it emerges that WALTzER will be able to constrain the lower and upper atmosphere for 54 exoplanets over its three years nominal mission lifetime. If instead we focus on single atmospheric tracers, the number of targets increases to 84 when considering the Na{\sc i} D lines, to 101 when considering the detection and characterisation of aerosols, and to 160 when considering the detection of planetary material outside of the Roche lobe (see Figure~\ref{fig:targets_differentN} and Section~\ref{sec:requirements}). 


\subsubsection{Complementarity with JWST and Ariel}
\label{sec:complementarity}

Of the 54 WALTzER prime targets, 53 are also part of the Ariel sample \citep{EdwardsTinetti2022ajArielTargetList}, and 22 have been, or are currently scheduled to be, observed with JWST (as of Cycle 5).
The complementarity of spectroscopic coverage in the NIR by Ariel or JWST and in the NUV/VIS by WALTzER creates extensive synergies between these missions.  An extended spectral coverage from the NUV to NIR will break degeneracies in the atmospheric modelling analysis, enabling us to disentangle the impact of atmospheric composition, cloud properties, escape, and stellar activity \citep[e.g.][]{RackhamEtal2018apjStellarHeterogeneityI, chakraborty2024, moran2023}. With a targeted launch in mid-2030, WALTzER would complement the observations from JWST and the large statistical survey planned by Ariel, which is currently scheduled for launch in 2031. WALTzER's flexibility also enables simultaneous observations with JWST and Ariel, removing ambiguities caused by temporal variability.

\begin{figure*}
\includegraphics[width=\textwidth]{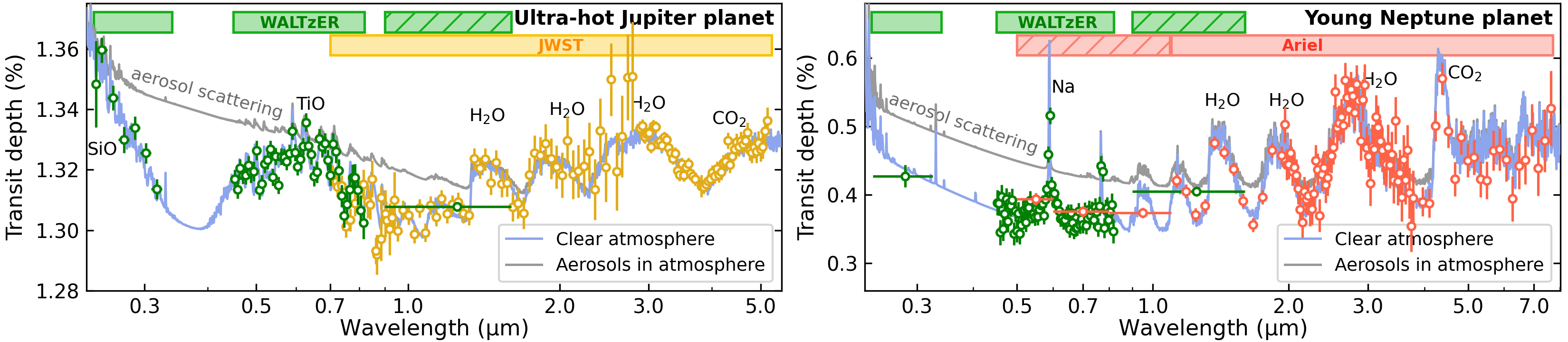}
\includegraphics[width=0.49\textwidth]{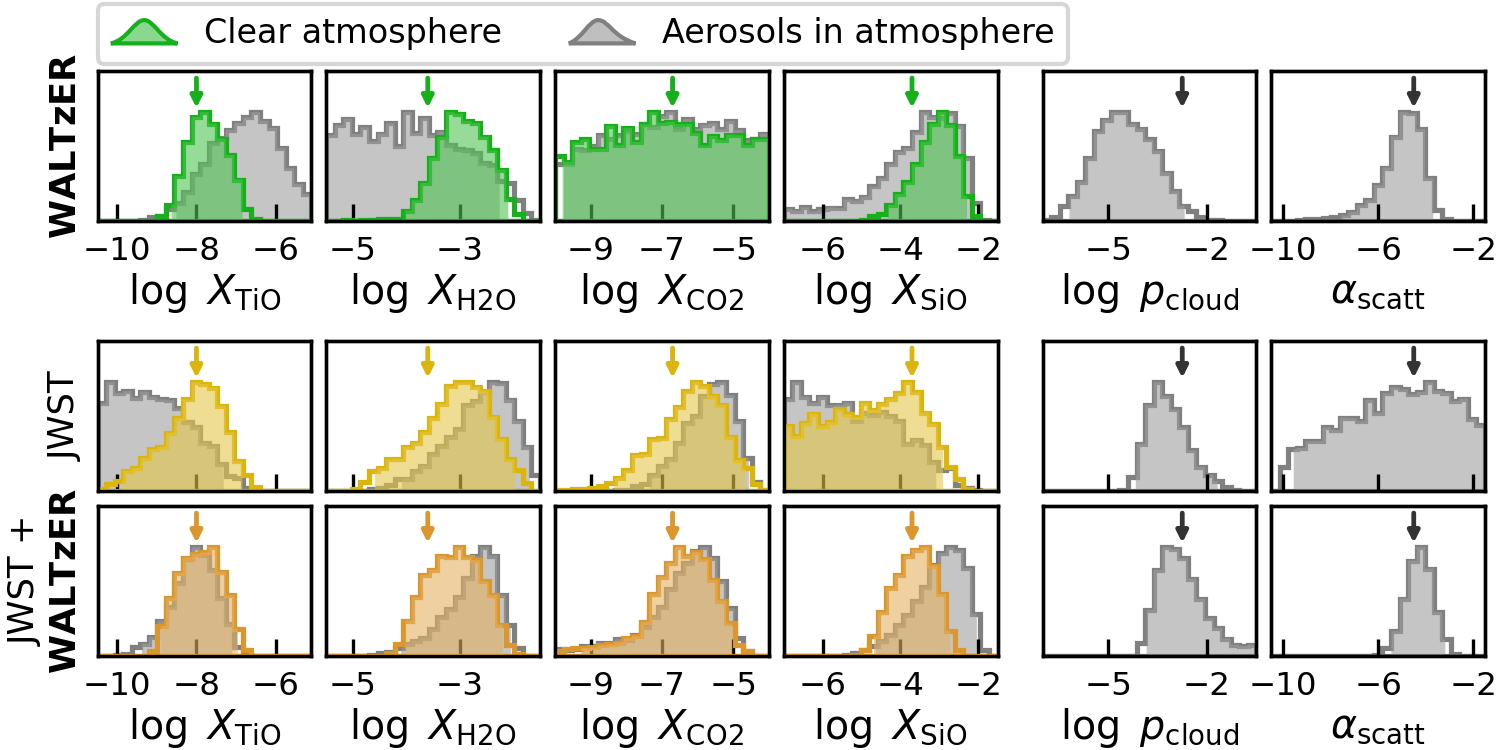}
\hfill
\includegraphics[width=0.49\textwidth]{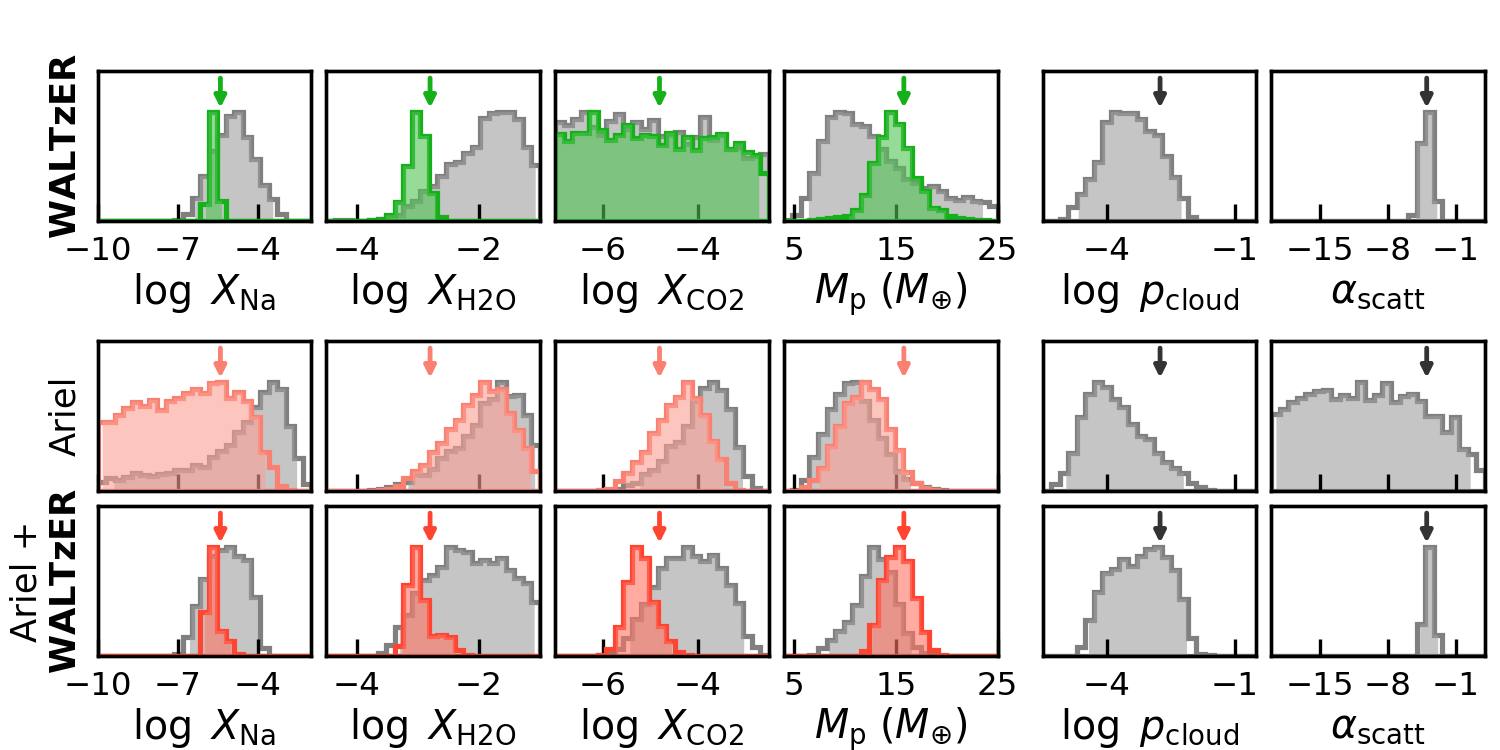}
\caption{Exoplanet atmospheric characterisation with WALTzER and in synergy with Ariel and JWST. {\bf Top:} transmission spectra of an ultra-hot Jupiter based on KELT-20b (left) and a young Neptune planet based on HIP\,67522b (right), assuming atmospheres with (gray) and without (blue) aerosols. The coloured markers with error bars show simulated observations with WALTzER (20 transits, green), JWST (single transit with NIRISS/SOSS and NIRSpec/G395H, gold), and Ariel (1 transit, red). The solid and hatched bars show the respective spectroscopic and photometric coverage. {\bf Bottom:} retrieval constraints obtained from WALTzER alone (top), by JWST or Ariel (middle), and in combination between WALTzER and JWST/Ariel (bottom). The coloured and grey histograms show the clear and aerosol-laden atmosphere cases, respectively, for selected key abundance species, planet mass ($M_{\rm p}$), aerosol cloud top pressure $p_{\rm cloud}$ (bar), and scattering power-law slope $\alpha_{\rm scatt}$. The arrows above each panel mark the true value of the respective parameter.}
\label{fig:synergy}
\end{figure*}

To characterize the constraining power of WALTzER, we employed the open-source \textsc{Pyrat Bay} modelling framework \citep{cubillos2021,CubillosEtal2026mnrasPyratBay2} to simulate exoplanet atmospheres, generate their transmission spectra, and perform Bayesian atmospheric retrievals based on WALTzER observations.
The atmospheric models consist of a set of parametrised temperature and composition profiles as a function of pressure in the 100--10$^{-9}$ bar range under hydrostatic equilibrium. We adopted an isothermal temperature profile and a free-chemistry abundance prescription, where the composition is parametrised by constant-with-altitude volume-mixing ratios (VMRs) of the species of interest.
The radiative transfer was computed at a spectral resolution of $R= 25000$ including line-sampled cross-sections for CO \citep{LiEtal2015apjsCOlineList}, \ch{CO2} \citep{HuangEtal2023jmospAmesCO2}, \ch{H2O} \citep{PolyanskyEtal2018mnrasPOKAZATELexomolH2O}, \ch{CH4} \citep{YurchenkoEtal2024mnrasExomolCH4mm}, \ch{H2S} \citep{AzzamEtal2016mnrasExoMolH2S}, \ch{SO2} \citep{UnderwoodEtal2016mnrasSO2exoamesExomol}, \ch{NH3} \citep{ColesEtal2019mnrasNH3coyuteExomol, YurchenkoEtal2024mnrasExomolNH3coyute}, SiO \citep{YurchenkoEtal2022mnrasSIOUVENIRexomolSiO}, TiO \citep{McKemmishEtal2019mnrasTOTOexomolTiO}, VO \citep{BowesmanEtal2024ExomolVOhyvo}, Na \citep{AllardEtal2019aaSodiumProfile}, and K \citep{AllardEtal2016aaPotassiumLine}; collision-induced absorption for \ch{H2}--\ch{H2} and \ch{H2}--He \citep{BorysowEtal1988apjH2HeRT, BorysowEtal1989apjH2HeRVRT, BorysowEtal2001jqsrtH2H2highT, BorysowFrommhold1989apjH2HeOvertones, Borysow2002jqsrtH2H2lowT}; and Rayleigh-scattering for \ch{H2} and He \citep{Kurucz1970saorsAtlas}.
Additionally, we modelled aerosol opacity via an opaque cloud-deck located at a given pressure level, and Rayleigh-like absorption parametrised by a cross-section scaling factor and scattering power-law slope \citep{LecavelierEtal2008aaRayleighHD189733b}.

We performed simulations for two planets representative of the WALTzER target sample, an ultra-hot Jupiter (similar to KELT-20b) and a young Neptune planet with an extended atmosphere (similar to HIP\,67522b). We generated transmission spectra for an atmosphere with VMRs consistent with a solar metallicity at the planet's equilibrium temperature. For each target we considered two scenarios, a cloud-free case and one with significant aerosol coverage (Figure~\ref{fig:synergy}).
For each model, we simulated WALTzER time-series observation spectra based on the WALTzER ETC noise estimations (see Section~\ref{section:etc}) at the end of the WALTzER nominal mission (i.e. 20 transits).
We then performed atmospheric retrieval analyses using nested-sampling posterior sampling \citep{Skilling2004aipcNestedSampling, Skilling2006baNestedSampling} with MultiNest \citep{FerozEtal2009mnrasMultiNest, BuchnerEtal2014aaBayesianXrayAGN}. The retrievals include 15--16 free parameters that fit for the temperature, planet mass (for HIP\,67522b), reference pressure at $R_{\rm planet}$, VMRs of the species listed above, and cloud parameters. To study the synergy with other missions, we also simulated observations from JWST for KELT-20b and Ariel for HIP\,67522b. The JWST noise calculation considers single-transit observations with NIRISS/SOSS and NIRSpec/G395H computed with \textsc{Gen TSO} \citep{Cubillos2024}. The Ariel noise calculation considers an Ariel tier 3 target (1 transit) computed with ArielRad \citep{MugnaiEtal2020exaArielRad}.

Figure~\ref{fig:synergy} shows that, while JWST and Ariel better constrain the signatures and abundances of NIR species such as \ch{H2O}, \ch{CH4}, and \ch{CO2}, WALTzER is better suited to constrain optical absorbers such as Na, TiO, and aerosols. In this synergy, WALTzER's broadband NIR measurement overlaps the spectroscopic bands of JWST and Ariel and is therefore essential for anchoring and combining these datasets. The bottom panels of Figure~\ref{fig:synergy} compare marginal posterior distributions for selected atmospheric parameters obtained from retrievals using WALTzER alone, JWST or Ariel alone, and the combined WALTzER+JWST/Ariel observations. The substantially narrower posteriors in the combined retrievals, particularly in cases where single-instrument posteriors were broad or unconstrained, demonstrate the benefit of coordinated observations between WALTzER and the NIR missions.

Finally, we emphasize the importance of a broad-wavelength spectroscopic coverage for the characterization exoplanet atmospheres. A wide spectral coverage not only probes the signatures of multiple atmospheric absorbers \citep[including aerosol compositions; e.g.,][]{GrantEtal2023apjWASP17bQuartzClouds}, but it also helps disentangle the contaminating effect of stellar activity, which is particularly pronounced for young and late-type (e.g., M and K) stars. In transmission spectroscopy, stellar spots and faculae induce wavelength-dependent transit light source variations \citep[TLS,][]{RackhamEtal2018apjStellarHeterogeneityI} that become increasingly larger toward shorter wavelengths \citep[e.g.,][]{moran2023}.  Therefore, the NUV and VIS channels of WALTzER are ideally suited to characterize stellar activity. Unocculted spots and faculae must be modelled and corrected simultaneously with the planetary characterization, and thus,
a broad spectral coverage helps mitigate degeneracies between atmospheric composition and nuisance parameters such as the TLS effect \citep[e.g.,][]{FairmanEtal2024ajOpticalCoverage, verma2025, Barstow2026arxivStellarNoise}, as it has been shown for young stellar systems \citep[e.g.,][]{MurphyEtal2026ajV1298cTransitJWST}.

\subsubsection{Organisation of the science observations} \label{sec:organisaiton}

WALTzER's core survey will require approximately 12,000 hours of observing time. Since many of these observations are time-constrained, they will naturally leave scheduling gaps that can be efficiently filled with predominantly non-time-constrained programmes, defined by the consortium (Section~\ref{sec:additional_consortium_surveys}) and by the community, for example through yearly open calls (Section~\ref{sec:community_science}). This operational model fully exploits WALTzER's design as a panchromatic observatory. Its broad wavelength coverage, together with its flexible pointing and scheduling capabilities, enables a diverse scientific programme extending well beyond the core survey, encompassing Solar System science, stellar astrophysics, and Galactic astronomy.

\subsection{Additional consortium surveys} \label{sec:additional_consortium_surveys}

Exploiting the scheduling gaps of the core survey, the WALTzER consortium will conduct time-flexible surveys ($\sim$21\% of the total time), tailored to the 2030s landscape: PLATO \citep{rauer2025} will have discovered potentially habitable planets, the Habitable Worlds Observatory \citep[HWO;][]{hwo_feinberg2024} and and the Large Interferometer For Exoplanets \citep[LIFE;][]{life_quanz2022} will be taking shape, ESA's Jupiter Icy Moon Explorer \citep[Juice;][]{juice_grosset2013} will be exploring Jupiter, its icy moons and magnetosphere, and Comet Interceptor \citep{comet_interceptor_jones2024} will be flying by a dynamically new comet. The surveys will connect WALTzER to these impactful missions, leveraging its unique simultaneous multi-band capability.

\subsubsection{Radiation Environments of Exoplanets}

XUV radiation, including stellar flares, is the main driver of atmospheric erosion \citep{lammer2003}, controlling the efficiency and timescale of mass loss. NUV and visible radiation profoundly shapes planetary atmospheres through heating and molecular dissociation. Flares add to this by exposing the atmosphere to frequent energetic radiation bursts, leading to chemical changes and episodic mass loss \citep{chadney2017}. Precise stellar spectra, particularly in the UV, are inputs for planetary atmospheric models. To study this connection at a population level, WALTzER will collect single NUV and visible stellar spectra for a sample of $\sim$500 planet-hosting stars, ranging well beyond the core survey parameter space in terms of planetary masses, ages, and stellar types. For a sub-sample of $\sim$50 targets, repeated observations will also determine flare rates and energies using spectral reconstruction via activity-sensitive lines such as Mg{\sc ii} h\&k, H$\alpha$, and H$\beta$ \citep[e.g.][]{hawley1991,montes2000}. Furthermore, WALTzER observations of PLATO targets will connect to the extensive radial velocity follow-up effort for these objects, joining NUV activity indicators (Mg{\sc ii} h\&k) with their optical counterparts \citep[e.g. Ca{\sc ii} H\&K;][]{buccino2008}.

The radiation environment of exoplanets is also an important factor in the context of habitability and the emergence of life. UV stellar irradiation is critical for the atmospheric evolution and photochemistry of terrestrial exoplanets and helps to constrain theories for the emergence of life. NUV radiation in the  200--300 nm band drives the abiogenic synthesis of RNA \citep[e.g.][]{powner2009,ritson2012,ranjan2016}. WALTzER will systematically observe this band for host stars of habitable zone planets. One hour of total integration time allows full-resolution NUV characterisation for G and K stars within 100 and 50 pc, respectively. M-dwarf hosts, including the closest system hosting a terrestrial planet, Proxima Centauri, become accessible with spectral binning to $R=100$. This easily exceeds $R>35$, required to reliably model the reaction rates of the photochemical steps in prebiotic synthesis \citep{ranjan2016}. WALTzER will characterise the radiative environments around {\rm i}) potentially habitable planets to be found by PLATO, {\rm ii}) non-transiting habitable-zone planets, and {\rm iii}) primary sample targets of the LIFE \citep{quanz2022} and HWO \citep{tuchow2025} missions. Longer-term monitoring constrains temporally varying stellar spectral energy distributions (SEDs), providing important context for potential biosignatures and surface UV environments on habitable-zone exoplanets \citep{Ridgway2023,Chen2021}. 

\subsubsection{Active bodies in the Solar System}

WALTzER has also applications for the study of Solar System objects, such as comets and the environments of giant planets and their moons. Comets are some of the most pristine remnants of the early Solar System, preserving information about volatile chemistry in the protoplanetary disk and the delivery of water and organics to terrestrial planets. However, major questions remain about the mechanisms driving cometary activity and compositional evolution. Particularly valuable are long-period comets (LPCs) originating from the distant Oort Cloud. Their volatile inventories are largely unaltered by solar heating, thus revealing the composition of the early solar nebula. An LPC will be visited by ESA’s Comet Interceptor in the early 2030s. Another uniquely informative class are interstellar objects (ISOs), formed in other stellar systems and entering our Solar System. ISOs such as 1I/`Oumuamua \citep{meech2017}, 2I/Borisov \citep{guzik2020}, and 3I/Atlas \citep{denneau2025} provide precious opportunities to study the diversity of protoplanetary disk chemistry and planetary system formation across the Galaxy. The forthcoming Vera C. Rubin Observatory Legacy Survey of Space and Time will greatly increase LPC and ISO discoveries \citep{seligman2018}. WALTzER will monitor the comae of $\sim$30 comets in the NUV and visible, tracing water (via OH), hydrocarbons (via C$_2$), and other O-bearing species. WALTzER observations will probe sublimation and thermal evolution in cometary nuclei, map activity and composition along their orbits, and disentangle intrinsic compositional differences from temperature-driven outgassing.

The magnetospheres of Jupiter and Saturn are populated by neutral gas and plasma supplied by geophysically active moons. At Jupiter, the moon Io delivers $\sim$1 ton\,s$^{-1}$ of initially neutral, later ionised O and S through sublimation and volcanic activity \citep{roth2025}. Na and K are often used as tracers for this mass transport \citep{yoneda2025}. Although usually stable, transient enhancements in gas and plasma emissions have been observed, but their origin remains unclear \citep{yoshioka2017,roth2025}. WALTzER will map and monitor emission from S{\sc ii}, O{\sc ii}, O{\sc i}, Na{\sc i}, K{\sc i}, and other species in the Jovian system complementing ESA’s Juice and NASA’s Europa Clipper \citep{europa-clipper_pappalardo2024} close-range and in-situ observations. At Saturn, Enceladus injects $\sim$300 kg\,s$^{-1}$ of ocean water into the magnetosphere through south-polar plumes \citep{postberg2018}. Much of the ejected material remains neutral, forming extended clouds of H$_2$O \citep{villanueva2023}, OH \citep{shemansky1993}, and other species, ultimately feeding Saturn’s E ring. Over the years 2005-2016, Cassini observations hinted a gradual decline in plume activity, which could arise from vent clogging \citep{denny2024}. WALTzER will monitor Enceladus’ plumes through NUV OH and visible Na and K observations \citep{zolotov2007}, constraining long-term plume variability. These measurements will help clarify the processes driving Enceladus’ activity and support future exploration, including ESA’s planning of the Enceladus orbiter and South polar lander \citep{martins2025}.

\subsection{Possible community science}\label{sec:community_science}

During the nominal mission, 25\% of WALTzER’s observing time will be made available to the community, for example through an open-time programme managed by ESA. There is a broad range of science cases that can be addressed with WALTzER. We give a few examples below.

\subsubsection{High-energy stellar phenomena}

Magnetic activity, flares, accretion, and binary interactions all produce variability on minute-to-hour timescales and couple emission from hot plasma to cooler stellar and circumstellar regions. WALTzER’s strictly simultaneous wavelength coverage removes degeneracies introduced by variability, enabling direct physical connections between emitting regions. The NUV traces the hottest components, including chromospheres and accretion shocks \citep[e.g. Mg{\sc ii} h\&k;][]{hawley1991}, while the visible and NIR capture the dynamical and cooler response. This enables a wide range of community-driven investigations across stellar populations. For example, simultaneous NUV-visible-NIR observations would characterise stellar magnetic activity and flare energetics across spectral types and ages \citep{wright2011,cardini2007}, linking NUV diagnostics to visible and NIR variability. In interacting binaries, this capability allows accretion flows and reprocessed emission to be traced consistently across a system, providing new constraints on accretion physics and geometry \citep{mukai2017,ritter2003,arkas2019}. 

White dwarf–main sequence (WDMS) binaries are powerful probes of star formation and stellar evolution. Their white dwarf (WD) cooling rates provide precise ages, enabling studies of age-dependent properties, such as activity, rotation, and metallicity, in companion stars \citep[e.g.][]{Rebassa-Mansergas2021,Rebassa-Mansergas2023,Raddi2022,Chiti2024}. Of particular interest are systems hosting a WD and a main-sequence (MS) AFGK-type star, progenitors of double WDs that can produce Type Ia supernovae (SN) and are major sources of low-frequency gravitational waves in the Galaxy, especially in the millihertz range targeted by the ESA Laser Interferometer Space Antenna \citep[LISA;][]{AmaroSeoane2023:LISA} mission. Their SEDs are composite: WDs dominate the ultraviolet emission, while the MS stars emit mainly in the visible and NIR. Combining WALTzER NUV spectra with Gaia parallaxes yields precise WD masses and temperatures, as well as system ages to $\sim$5\% precision, while visible spectra and NIR photometry constrain the properties of the companion. In WDMS binaries, the MS star outshines the WD in the optical, making ultraviolet observations essential for white-dwarf detection \citep[e.g.][]{parsons2016}. WALTzER NUV spectra could identify thousands of such systems, constraining binary evolution, common envelope efficiency, and the formation of gravitational wave sources and SN Ia progenitors.

However, while 95\% of all stars will eventually become WDs, their paths to this endpoint are diverse. This is most evident in the population of hot subdwarf stars: mostly helium-core stars that lack any hydrogen envelope. They occupy the Hertzsprung-Russell diagram region between the MS and WD, covering a vast temperature (10kK--100kK) and luminosity range (1--10$^4$ times solar). Besides canonical post-asymptotic giant branch (AGB) evolution, they are formed by close binary evolution with mass transfer, common envelope episodes, and mergers. Their peculiar chemical compositions, magnetism, and stellar winds allow us to test nucleosynthesis and atomic diffusion calculations. Several subdwarf binaries have orbital periods shorter than 1 hour and these binaries contribute a small but significant fraction of type Ia SNe, particularly important for recovering measurements of short type Ia delay times. Furthermore, they qualify as calibration sources for gravitational wave detectors such as LISA. Links between the subclasses of hot subdwarfs can only be understood by detailed far-UV abundance studies, because most metals become accessible only through their UV lines. WALTzER NUV spectroscopy could allow one to $(i)$ identify stars formed by merger channels, as well as stable and unstable Roche lobe overflow evolution, and thus constrain binary evolution in general; $(ii)$ obtain observational constraints to test theoretical nucleosynthesis, atomic diffusion, stellar wind, and evolutionary model calculations; $(iii)$ identify hitherto undetected trans-Fe elements, study the interstellar medium, and provide well-characterised stars as new spectrophotometric standards. Given the currently known space density \citep{dawson2024}, we expect a few hundred of these objects with a UV magnitude brighter than 14 mag, and thus reachable by WALTzER.

\subsubsection{Confirming nearby habitable-zone rocky worlds}

Community time could also be used for exoplanet science not covered by the core survey, e.g. to confirm planet candidates. Population estimates suggest that several tens of transiting planets should exist in the habitable zones of M dwarfs within 30 pc, but just a few are known. Many remain undetected because their transits are shallow and often fall near the sensitivity limits of wide-field surveys \citep[e.g. TESS;][]{ricker2015}. Ground-based validation is further limited by atmospheric scintillation and day–night cycles, which restricts follow-up to specific observing windows. WALTzER’s high-precision NIR photometry can confirm (or reject) these candidate planets, by systematically following up transit signals at and slightly below the common threshold of 7.1\,$\sigma$ \citep{jenkins2002}. When observing M dwarfs, WALTzER‘s larger collecting area and NIR detector will significantly outperform CHEOPS and PLATO (in areas covered with $\leq$12 cameras), and its near all-sky accessibility provides strong synergy with PLATO by enabling follow-up and validation across the most of the celestial sphere. The simultaneous visible spectroscopy gives an independent transit depth measurement and probes stellar activity features, such as star spots, as the planet sequentially occults different regions of the stellar disk \citep[e.g.][]{sanchis2011}. This improves the robustness of planetary radius measurements and offers automatic multi-wavelength validation \citep{parviainen2019}. We estimate $\sim$1200 M dwarfs within 30 pc (V mag 6--18) are amenable to 3\,$\sigma$ transit detections of Earth-like planets in single visits with WALTzER.

\begin{figure}
	\includegraphics[width=\columnwidth]{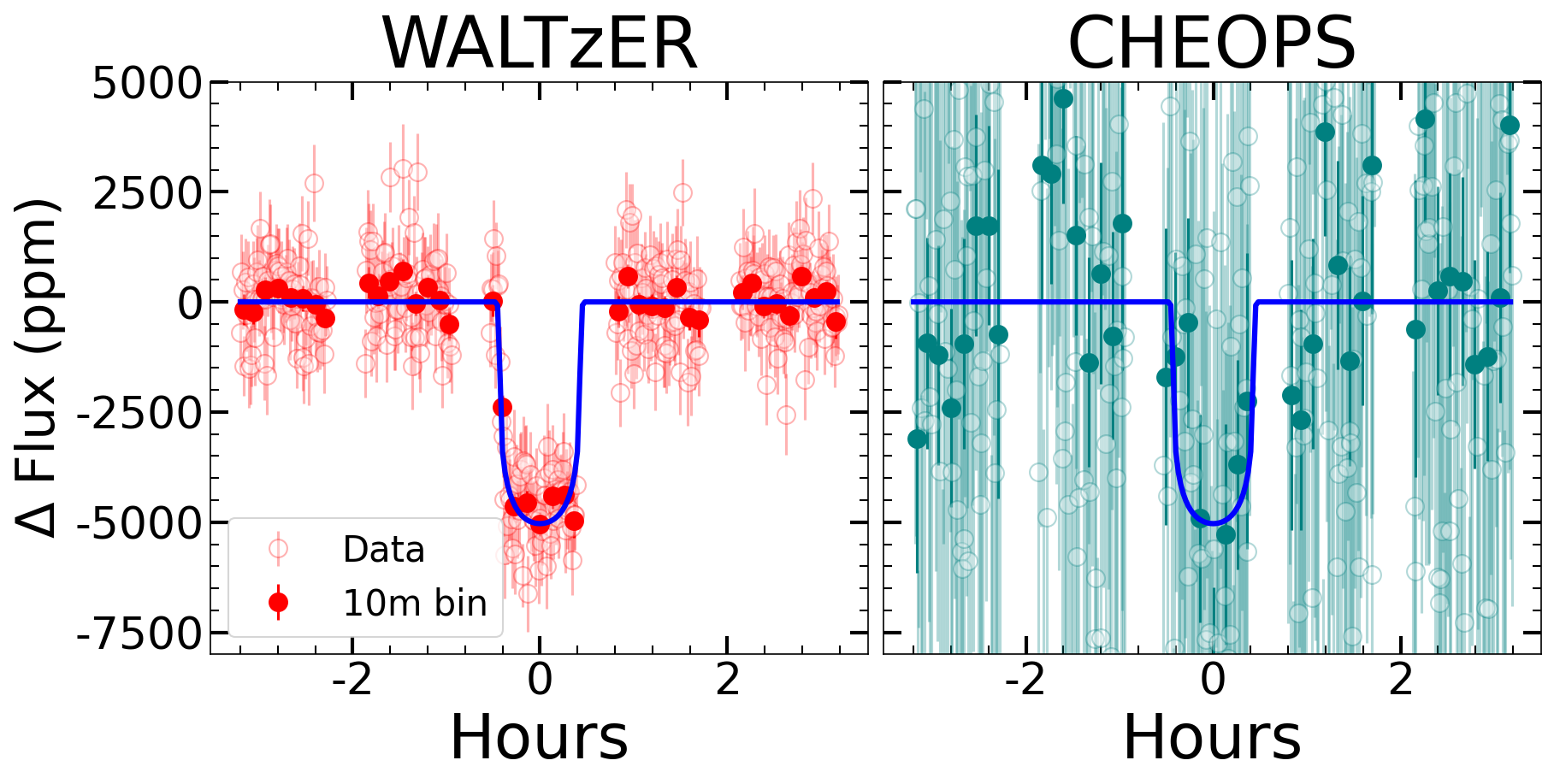}
    \caption{Simulated photometric data for the transit of an Earth-sized planet on V=16 mag M6V dwarf. The left panel shows WALTzER's NIR photometry with the transit model. The right panel shows the same for CHEOPS.}
    \label{fig:transit_comp}
\end{figure}

Figure \ref{fig:transit_comp} illustrates this capability by comparing simulated WALTzER and CHEOPS light curves for a shallow transit of an Earth-sized planet across a V=16 mag M6V dwarf. WALTzER’s NIR photometry recovers the transit with high significance from a single event, while the same signal in CHEOPS is buried in the noise. This demonstrates why WALTzER can validate low-S/N habitable-zone candidates that are inaccessible to current facilities. Using only $\sim$25\% of WALTzER’s idle time, a survey with this goal could validate up to 11 systems, a $>$75\% increase over the current sample. The full idle-time budget could yield up to 20 systems. These validated planets would be prime targets for atmospheric characterisation with JWST \citep{jwst_gardner2006}, ELT/ANDES \citep{andes_marconi2022}, and high-precision RV facilities \citep[ESPRESSO, NIRPS, MAROON-X, KPF;][]{espresso_pepe2021,nirps_bouchy2025,maroon-x_seifahrt2018,kpf_gibson2016}.

\section{Instrument and mission requirements}
\label{sec:requirements}

The core exoplanet science described above drives most of the requirements, which naturally lead to a versatile instrument capable of supporting a wide range of science cases. 

\subsection{Requirements}
\noindent{\bf Wavelength coverage and spectral resolution.} WALTzER shall simultaneously cover the NUV, visible, and NIR bands, where the instrument shall spectroscopically cover in the NUV Fe{\sc i} (248, 255, 272 nm), Fe{\sc ii} (259, 260 nm), Mg{\sc i} (285 nm), and Mg{\sc ii} (280 nm) features, in the optical the Na{\sc i} D (589 nm) and K{\sc i} (770 nm) doublets spanning at least a 300 nm wide band to constrain the continuum diagnosing aerosols, and in the NIR photometrically covers/overlaps with the water features in the 900--1600 nm band. 

\begin{figure}
	\includegraphics[width=\columnwidth]{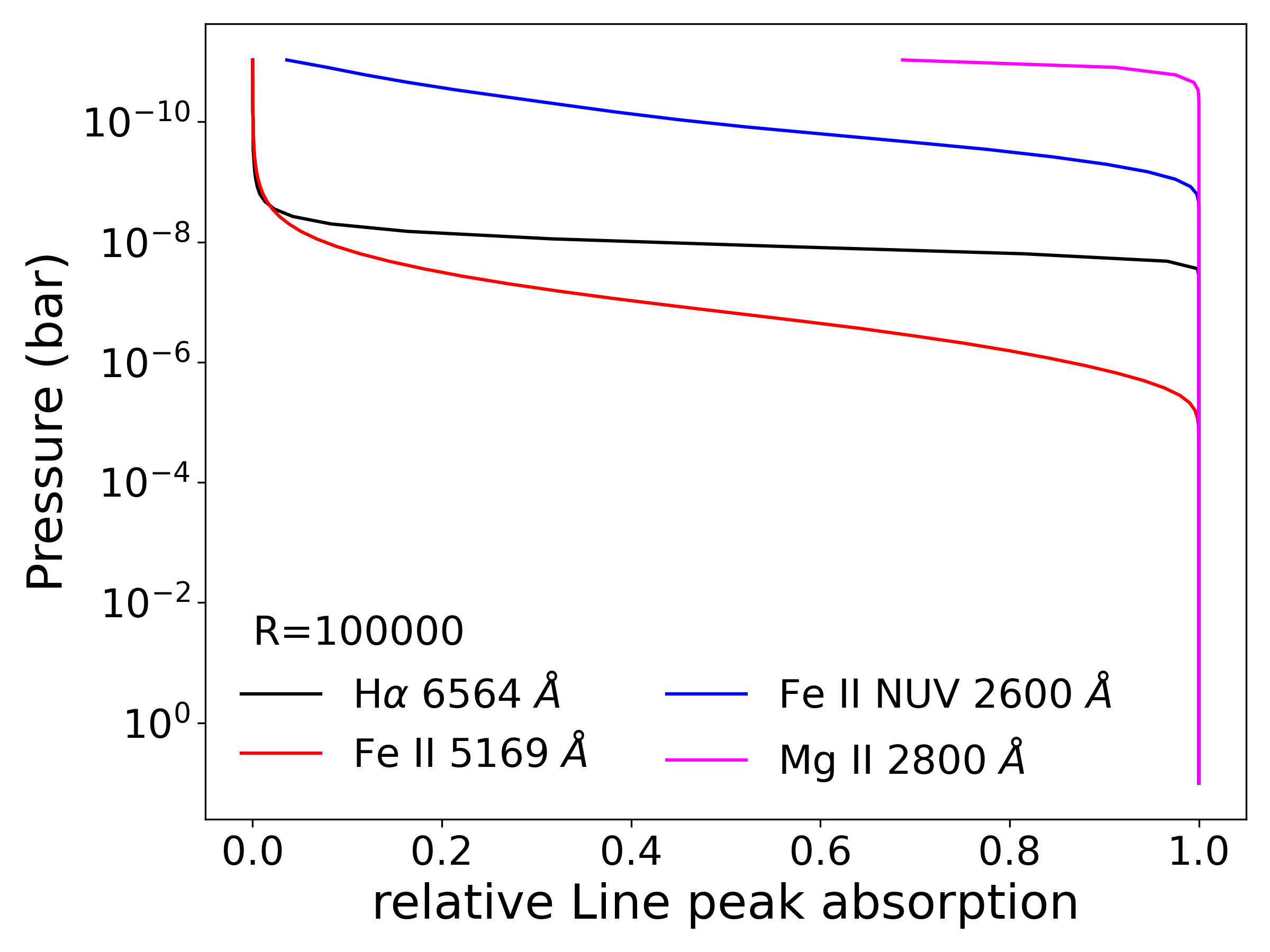}
	\includegraphics[width=\columnwidth]{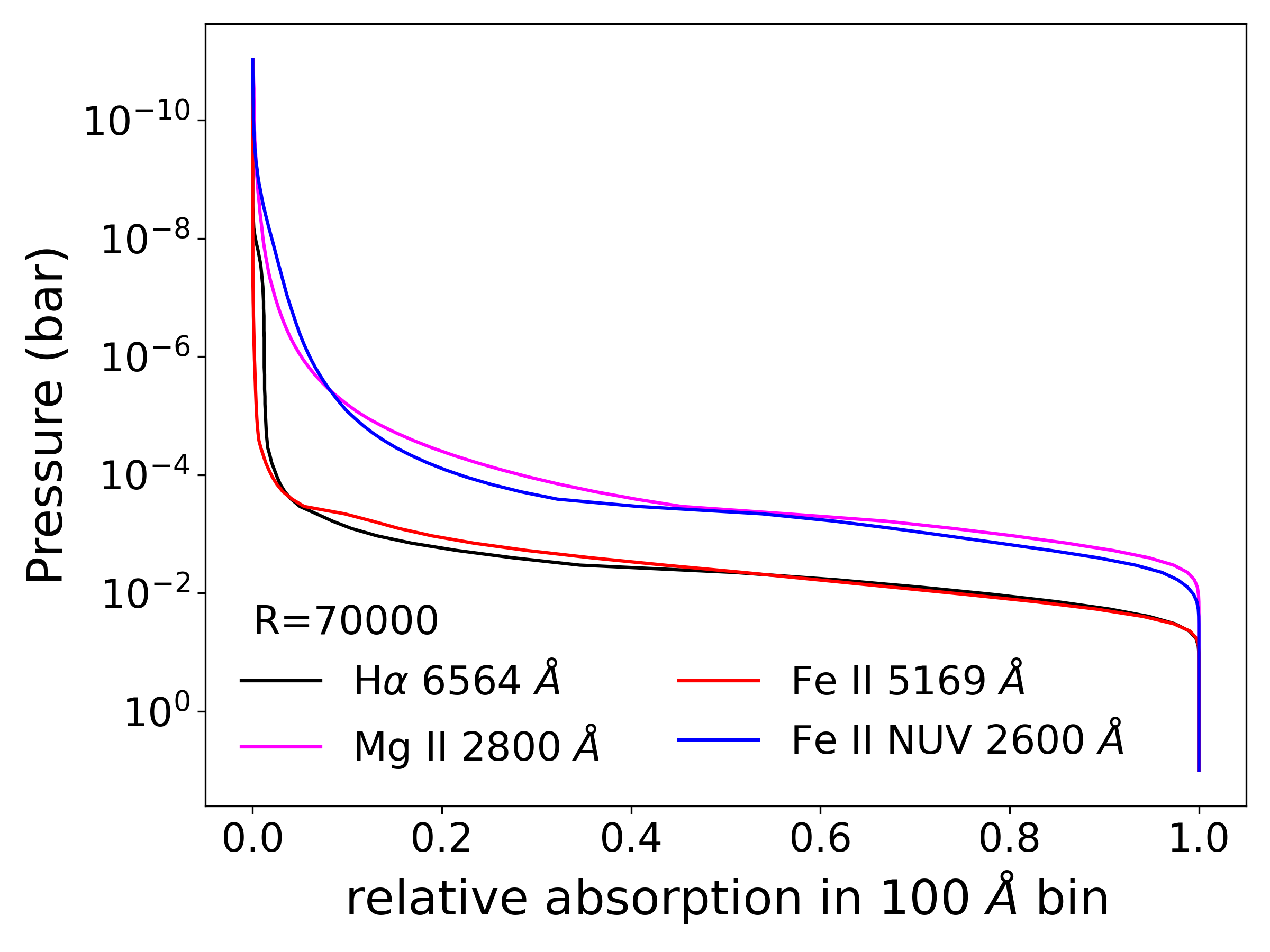}
    \includegraphics[width=\columnwidth]{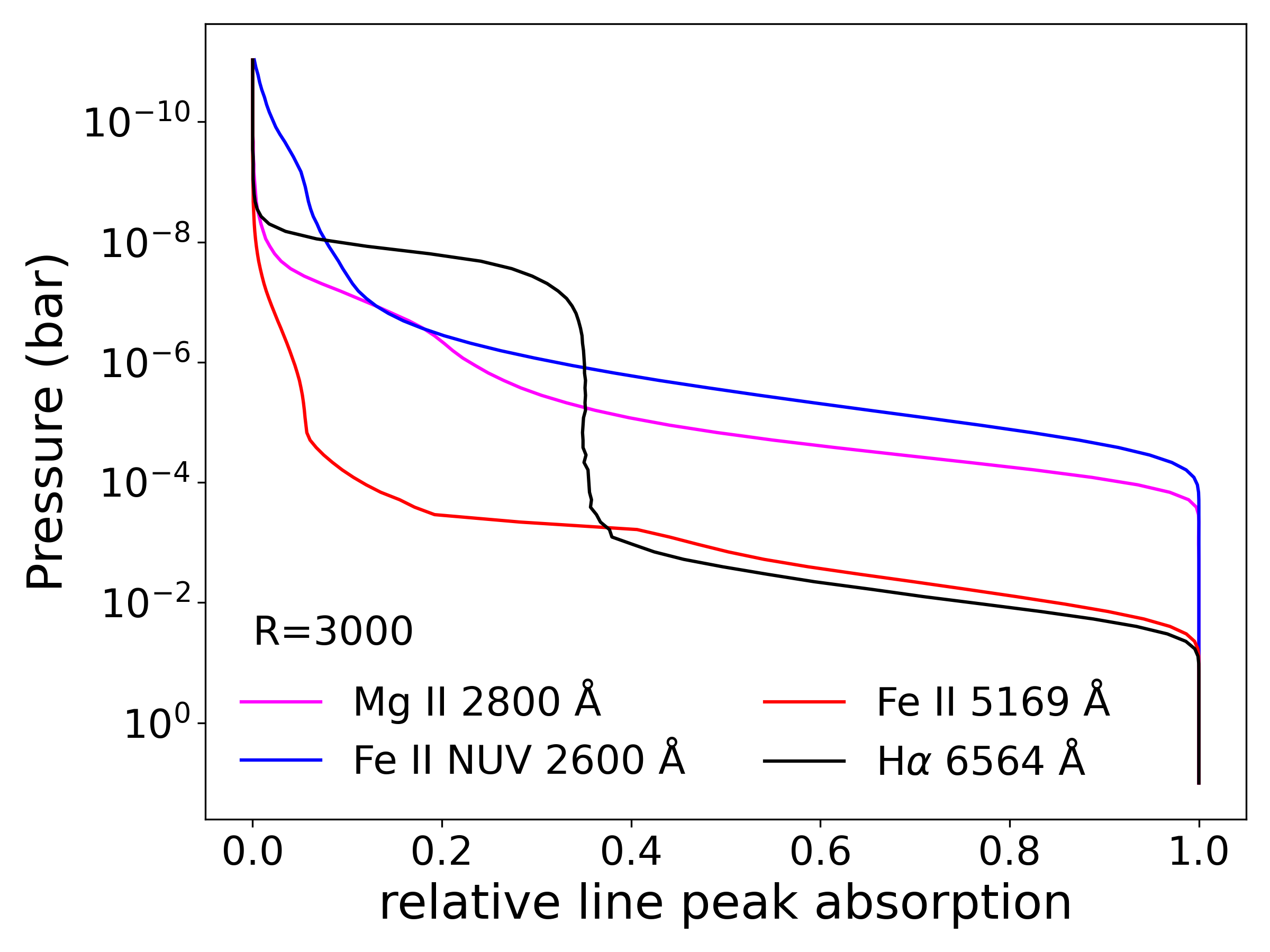}
    \caption{Relative peak absorption of representative NUV and visible features (H$\alpha$, black; Fe{\sc ii} at $\sim$5169\,\AA, blue; Mg{\sc ii} h\&k, red; Fe{\sc ii} at $\sim$2600\,\AA, magenta) as a function of atmospheric pressure for the UHJ WASP-178b at a resolving power of 100,000 (top), 70 (middle), and 3000 (bottom). The line formation region corresponds to the pressure range in which the relative line peak absorption lies between zero and one.}
    \label{fig:exoplanet_resolution}
\end{figure}

The spectral resolution requirement is driven by the need to resolve the Mg{\sc ii} h\&k and Na{\sc i} D stellar doublets, which calls for a resolving power $R=\lambda$/$\Delta\lambda$ $>$1500, with a goal of $R\approx$3000, and by the need to simultaneously probe both the lower and upper atmosphere of gas-rich exoplanets. Figure~\ref{fig:exoplanet_resolution} shows the line formation region of representative NUV and visible features for the UHJ WASP-178b computed at different resolving powers \citep[see][for more details]{Fossati2025}. At high resolution, the line cores probe mainly the top of the atmosphere and, because there are no stable high-resolution spectrographs in space, information about the location and shape of the continuum is lost during the analysis. For this reason, rebinning high-resolution observations does not lead us to gain information about aerosols in the lower atmosphere. At the resolution of HST/WFC3/UVIS, even the strongest spectral features probe only the lower atmosphere. Instead, at a resolution of 3000, combining visible and NUV spectral features enables one to probe all atmospheric layers at the same time and the space-based location enables one retaining critical continuum information.

\noindent{\bf Orbit and sky coverage.} WALTzER's target orbit is a 6 am – 6 pm dawn-dusk Low-Earth Orbit (LEO) at a 600--700 km altitude (baseline 700 km). To ensure that the radiators are never illuminated by the Sun and to minimise straylight, WALTzER implements a 110$^\circ$ Sun-exclusion angle, a 20$^\circ$--30$^\circ$ Earth limb exclusion angle (average 25$^\circ$; adaptable depending on the target magnitude), and a 5$^\circ$ Moon exclusion angle. With these parameters WALTzER covers $\sim$90\% of the sky for $\geq$10 days, maximising the number and observability of known gas-rich planets and enabling a significant overlap with PLATO.

\noindent{\bf Wavelength calibration and stability.} Spectral line identification requires the wavelength precision and accuracy over the maximum exposure time (goal: over a full transit observation) to be better than a resolution element (goal: half a resolution element). Wavelength stability to better than half a resolution element is required over a time scale equal to the maximum exposure time, which is set equal to 1 hour to limit the impact of cosmic rays.

\noindent{\bf Flux calibration and stability.} To match/compare WALTzER fluxes with those obtained by other space-based facilities (e.g. HST) we set a flux calibration requirement of $\leq$2\%, before converting WALTzER fluxes to physical fluxes using standard stars. WALTzER shall be capable to maintain its photometric performances over 40 hours (i.e. four times the transit duration of the planets in the current target list with the longest transit duration).

\noindent{\bf Bright-to-faint target magnitudes.} WALTzER shall observe stars with V-band magnitudes ranging between 6 and 12, with the required photometric stability. WALTzER shall also be able to observe stars ranging from magnitude V\,=\,$-$1.5 mag (i.e. Sirius) to V\,=\,16 mag (i.e. M dwarfs), but no photometric and pointing requirements are set for stars outside of the 6$\leq$V$\leq$12 mag range.


\noindent{\bf Pointing accuracy and guiding.} To ensure resolving power and flux stability WALTzER shall have a high pointing accuracy. Specifically, a centroid estimation error $<$0.44\arcsec~is required. This can be achieved by employing a pointing in the loop (PITL) system. WALTzER shall also be able to observe Solar System objects. Therefore, the satellite must be capable of pointing and guiding on them.

\noindent{\bf Mission duration.} WALTzER’s nominal mission lasts 3 years. This enables observing $\leq$20 transits each for $>$30 gas-rich planets, and thus study the short- (between consecutive transits), medium- (between transits in the same season), and long-term (between transits in different seasons) atmospheric variability for many of them. Fuel in the spacecraft shall be the only consumable.

\noindent{\bf Sensitivity.} Given the panchromatic nature of WALTzER, we adopted three criteria defining the sensitivity requirements for each channel and, if fulfilled, enabling us to achieve the core science goals: WALTzER shall be able to (1) detect Na absorption in a clear atmosphere at a significance $>$5$\sigma$, (2) distinguish clear from aerosol-laden atmospheres at $>$5$\sigma$, and (3) determine whether planets fill their Roche-lobe radius at $>$3$\sigma$, for a sample of $>$30 targets in $\leq$20 transits. Fulfilling these criteria requires a photometric precision of 8652 ppm at 3000\AA~  ($\Delta\lambda$=0.50\,\AA, NUV channel), 1263 ppm at 5500\AA~ ($\Delta\lambda$=0.92\,\AA, visible channel), and 23 ppm at 12500\AA~ ($\Delta\lambda$=700 nm, NIR channel), to be achieved in a 6-hour integration time of a V\,=\,10 mag G0V star.

Meeting these requirements also enables WALTzER to fulfil the other scientific goals and secondary surveys discussed above. To demonstrate that an instrument built with the sensitivity defined above fulfils the desired scientific requirements, we developed an exposure time calculator (ETC; see Section~\ref{section:etc}) based on the WALTzER design, which we applied to the known exoplanet population from the NASA exoplanet archive \citep{christiansen2025}, filtered by excluding targets with faint hosts (V\,$>$\,12 mag), small planetary radii (R\,$<$\,0.4 R$_{\rm J}$), and long-period orbits ($>$20 days). This leaves a pool of $\sim$350 transiting exoplanets to be considered.

\subsection{Assessments of instrument performance vs requirements}

\subsubsection{Exposure Time Calculator}
\label{section:etc}

To assess WALTzER's radiometric performance, we developed an open-source ETC based on the heritage of the JWST Pandeia ETC \citep{Pontoppidan2016} and its exoplanet time-series extension, \textsc{Gen TSO} \citep{Cubillos2024}. The ETC allows users to select or provide spectral energy distributions (SEDs) for point or extended sources, specify exposure times, and configure the instrument setup for time-series observations. For transit observations, it propagates uncertainties into the simulated transit spectrum from a user-supplied transmission spectrum. Detector count rates are computed using the effective area of each channel. The WALTzER ETC code is available under \href{https://github.com/pcubillos/waltzer_etc}{https://github.com/pcubillos/waltzer\_etc}.

\begin{figure*}
\includegraphics[width=0.85\linewidth,clip]{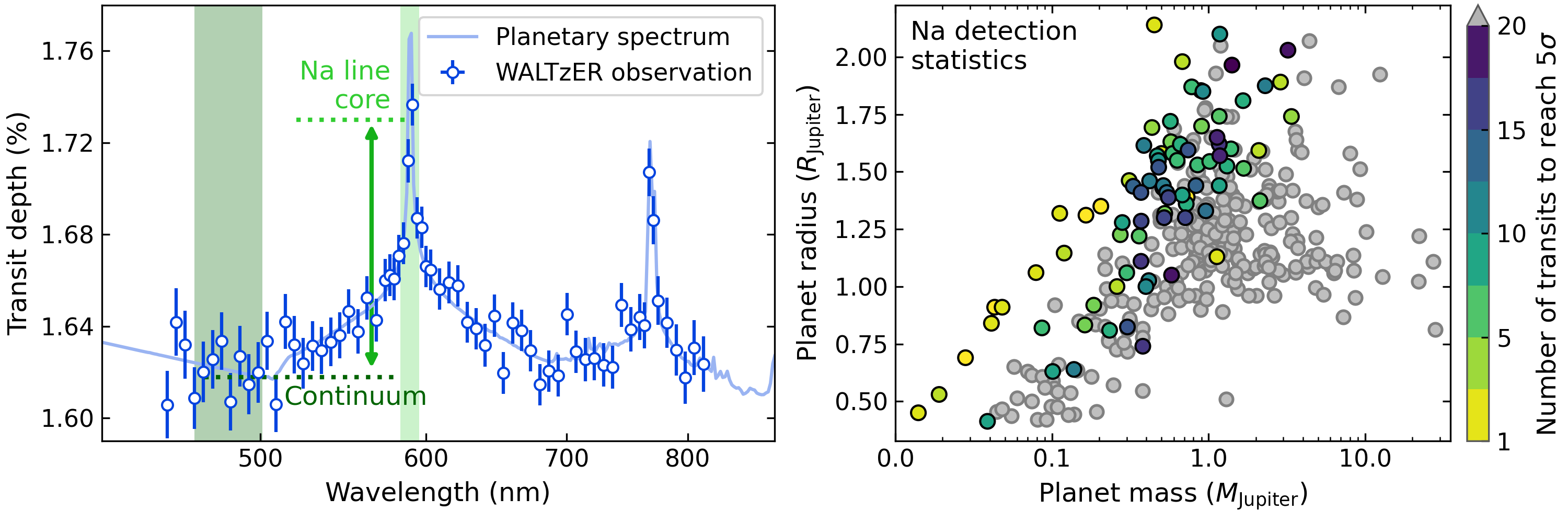}
\includegraphics[width=0.85\linewidth,clip]{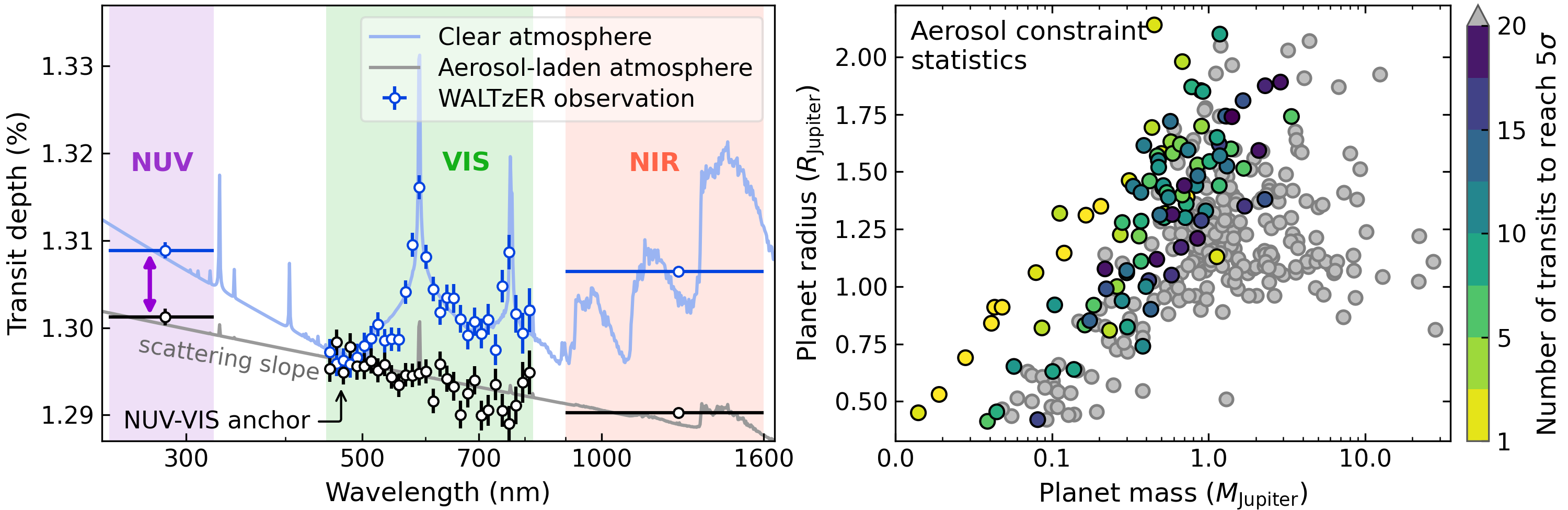}
\includegraphics[width=0.85\linewidth,clip]{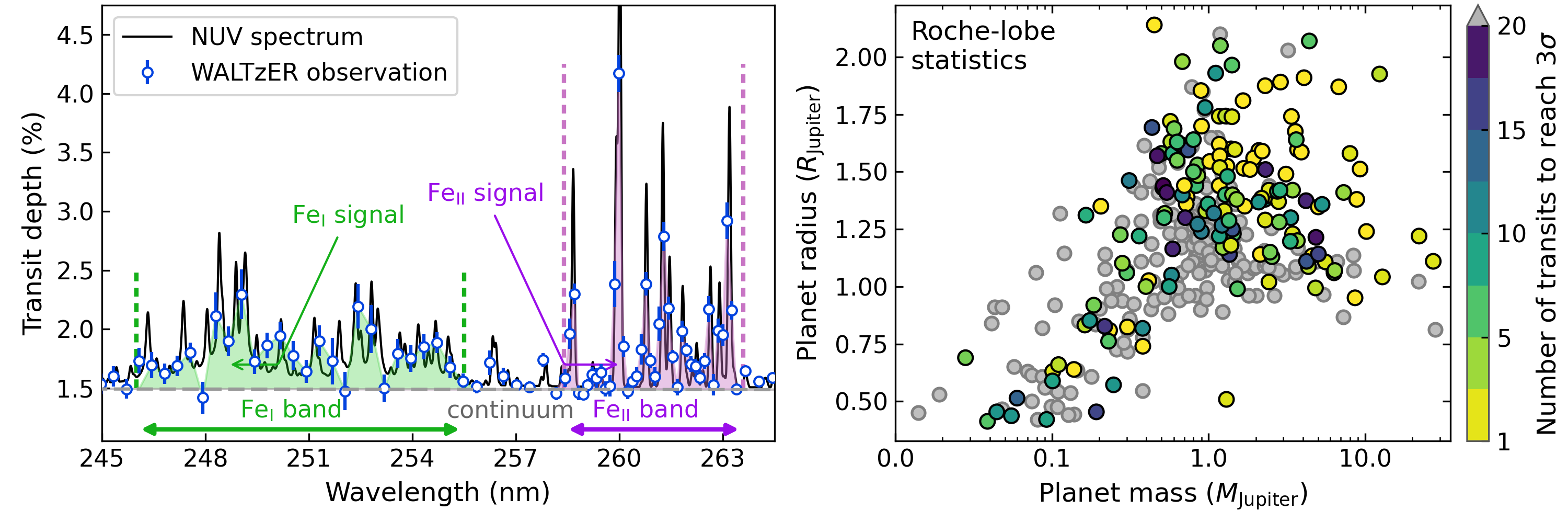}
\caption{WALTzER sensitivity requirements. {\bf Top:} for sodium detection we simulate transmission spectra of cloud-free atmospheres (blue curve) and expected WALTzER observation (blue markers with error bars). We quantify the Na-detection significance from the transit-depth difference between the core and nearby continuum of the Na line (light- and dark-green shaded areas, respectively). We expect WALTzER to detect Na for 84 exoplanets at $>5\sigma$ significance with $\leq$20 transits (right panel).
{\bf Middle:} for aerosol characterisation we simulate transmission spectra of atmospheres with and without aerosols (grey and blue, respectively) and expected WALTzER observations (circle markers). The aerosol-laden model has been vertically offset (matching the clear atmosphere at the labelled anchor point) to highlight the impact of aerosols on the scattering slope. We expect WALTzER to distinguish clear from aerosol-laden atmospheres for 101 exoplanets at $>5\sigma$ significance with $\leq$20 transits (right panel).
{\bf Bottom:} for the Roche-lobe filling characterisation we simulate NUV transmission spectra of planets filling their Roche-lobe radius (black curve) around Fe{\sc i} and Fe{\sc ii} bands (regions within green and purple dotted lines). Blue markers show a WALTzER observation. An Fe{\sc ii} signal is estimated by the area between the observed spectrum and the continuum inside the Fe{\sc ii} bands (purple shaded area). Similar calculations are done for Mg{\sc ii}. We expect WALTzER to detect the signature of Fe{\sc ii} or Mg{\sc ii} for 160 exoplanets at $>3\sigma$ significance with $\leq$20 transits, if the planet fills the Roche lobe (right panel).
}
\label{fig:requirements}
\end{figure*}

We assess the detectability of Na by examining its main feature in the visible channel ($\sim$590 nm; Figure \ref{fig:requirements}, top left). For each planet, we simulated the transmission spectrum assuming solar composition and a cloud-free atmosphere at the planet’s equilibrium temperature. We then used the ETC to simulate a WALTzER transit observation with its uncertainties. We approximated the Na detection significance by the Na signal-to-noise ratio (S/N), calculated from the difference between the transit depth in the core of the Na feature ($\sim$30\,\AA\ bin; light-green shaded area) and in the nearby continuum at $\sim$500 nm (dark-green shaded area), accounting for the uncertainties (Figure \ref{fig:requirements}, top left). We then determined the number of transits $N$ required for a Na detection at S/N$>$5 by scaling the single-transit S/N by $\sqrt{N}$. We identified 84 systems that meet this requirement in $\leq$20 transits (Figure \ref{fig:requirements}, top right) spanning the range in mass and radius of the core exoplanet survey.

We assess whether WALTzER can distinguish atmospheres with or without aerosols. First, we simulate aerosol-laden transmission spectra based on a kinetic, non-equilibrium cloud formation model \citep{helling2004,helling2023} applied on a grid of global circulation models \citep{carone2026}. For each planet, we simulated a transmission spectrum by adopting the grid model closest to the system's parameters, scaling by the planetary pressure scale height and stellar radius accordingly. We then used the ETC to simulate a WALTzER transit observation with its uncertainties (Figure \ref{fig:requirements}, middle left). Finally, we employed atmospheric retrievals \citep{cubillos2021, CubillosEtal2026mnrasPyratBay2} to constrain the aerosol properties and assess whether the WALTzER observations robustly detect the presence of aerosols \citep[e.g.,][]{benneke2013}. In these retrieval models, aerosols are parametrised via a grey cloud deck and a haze scattering power-law model \citep{LecavelierEtal2008aaRayleighHD189733b}. The retrieval analyses indicate that WALTzER can distinguish clear from aerosol-laden atmospheres at $>$5$\sigma$ with $\leq$20 transits for 101 systems (Figure \ref{fig:requirements}, middle right). The constraint of the scattering slope deserves particular attention, as it constrains the aerosol nature and particle-size distribution. WALTzER can place constraints at S/N$>$3 on the scattering slope for $\sim$30 targets based on the visible channel. For a dozen targets, the NUV channel will provide additional spectroscopic constraints on the scattering slope, relative to the visible channel (Figure \ref{fig:requirements}, middle left).

To assess the feasibility of detecting whether planets fill their Roche Lobe, corresponding to an ongoing mass loss, we focused on the NUV channel. First, we generated a grid of atmospheric models using a 1D self-consistent hydrodynamic code assuming representative planetary and stellar properties of our target list \citep{kubyshkina2024}. For the hosts' SED, we used PHOENIX models \citep{husser2013} supplemented with MUSCLES NUV spectra \citep{france2016} scaled over a grid of magnitudes. For each grid system, we computed disk-integrated transmission spectra using the EvE code \citep{bourrier2015}. Then, for each target we scaled the models by the system properties and processed through the ETC to obtain the transmission spectra and uncertainties (Figure \ref{fig:requirements}, bottom left). We consider that a planet's Roche Lobe is detected if either Mg{\sc ii} or Fe{\sc ii} absorption is detected with an amplitude corresponding to the transit depth of the Roche Lobe, in excess of the planetary continuum. We thus compute the S/N of excess absorption measured over one transit in the transmission spectrum of a simulated planet filling its Roche Lobe and estimate the number of transits required to reach S/N\,=\,3. There are 160 systems for which WALTzER could detect Roche-lobe filling of Fe{\sc ii} or Mg{\sc ii} at S/N$>$3 with $\leq$20 transits (Figure \ref{fig:requirements}, bottom right).

%

\section{WALTzER Instrument}\label{sec:instrument}

\begin{figure}
	\includegraphics[width=\columnwidth]{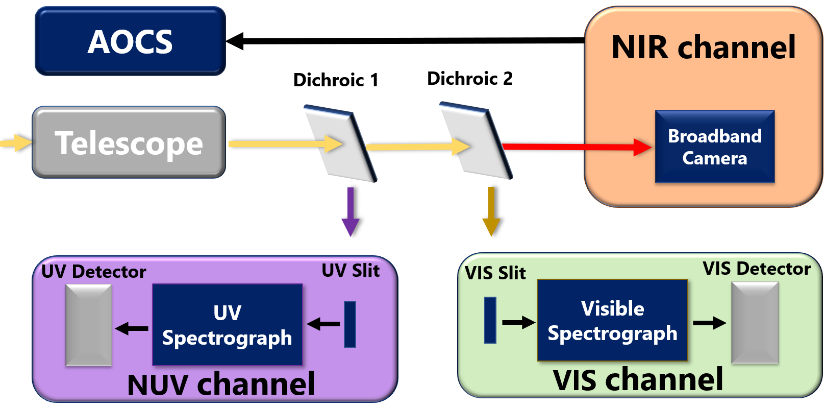}
    \caption{WALTzER block diagram.}
    \label{fig:block_dia}
\end{figure}

\begin{figure*}
	\includegraphics[width=0.9\textwidth]{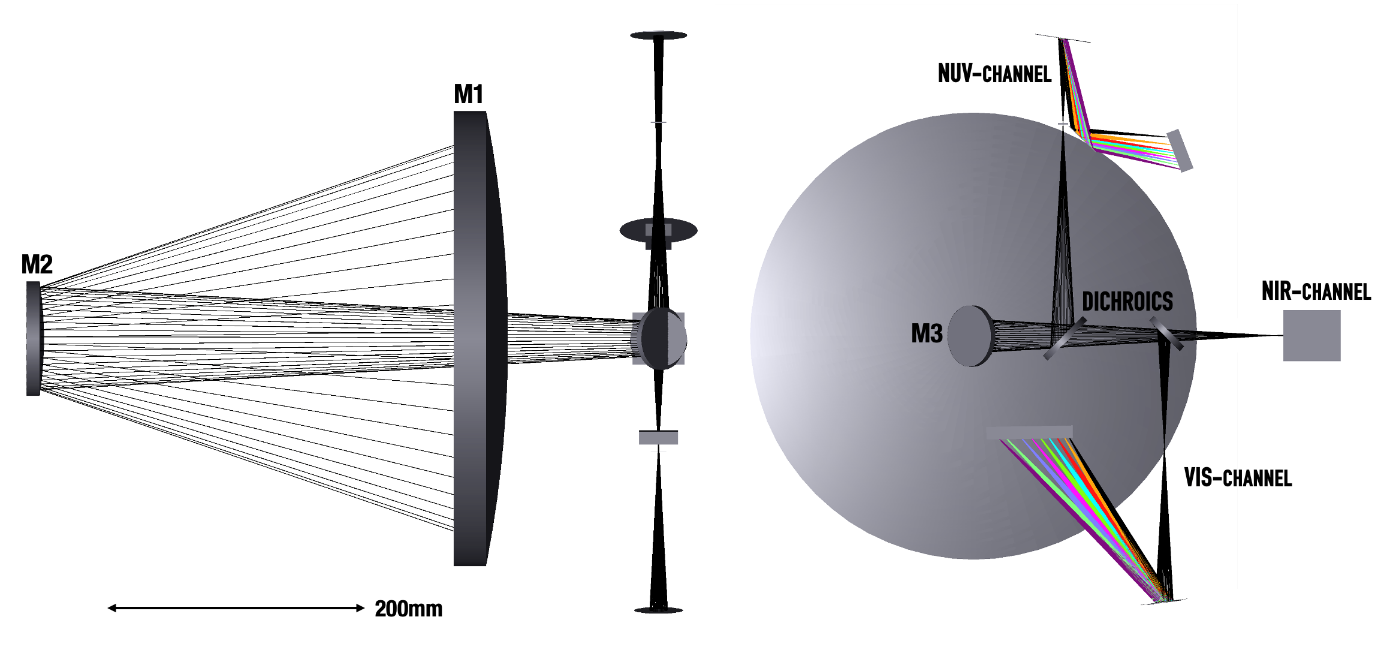}
    \caption{WALTzER optical layout. {\it Left}: optical ray paths in the $x$--$y$ plane, showing the telescope Primary (M1) and secondary (M2) mirrors. {\it Right}: corresponding ray paths projected onto the $y$--$z$ plane, illustrating the light distribution system and the three channels. The two projections together show the three-dimensional geometry and optical layout of the entire payload.}
    \label{fig:optical_dia}
\end{figure*}

WALTzER is a three channel instrument that simultaneously obtains medium resolution ($\sim$3000) NUV (240--325 nm) and visible (450-820 nm) spectra, and NIR (900-1700 nm) photometry. The instrument is fed by a 35 cm telescope. Figure~\ref{fig:block_dia} shows the instrument block diagram. The telescope feeds light into the three channels through a light distribution system (LDS) which consist of a fold mirror and two dichroics, enabling simultaneity. The platform provides coarse pointing with an error of 1\arcsec\ (relative pointing error, RPE), but an Attitude and Orbit Control System (AOCS) loop between the NIR camera and the platform allows refining of the pointing to an APE of $<$0.44\arcsec\ to avoid for example degradation of spectral resolution and slit losses (an important source of systematic noise). The NIR camera also provides simultaneous reference light curves (e.g. for timing), enables the correction of systematic noise, and supports the identification of stellar spot crossing events. Importantly it allows cross-calibration with JWST and Ariel. The payload is controlled by the main electronic box (EBOX), which provides power, drives the instrument, and formats the data. The Spacecraft Service Vehicle Module (SVM) provides power, pointing function, communication, and orbital manoeuvre capability (e.g. collision avoidance, final de-orbiting).

\subsection{Optical design}

The payload comprises a Cassegrain telescope featuring a 35 cm parabolic primary mirror (M1) and a hyperbolic secondary mirror (M2) with an F\# of $\sim$7.7. Both mirrors are made of Zerodur with protected Al+MgF$_2$ coating. The light is reflected by the fold mirror (M3), which then encounters a dichroic (D1) reflecting NUV light ($<$350 nm) into the NUV spectrograph. Longer wavelengths are transmitted and directed to a second dichroic (D2) that reflects visible light (350--835 nm) into the visible spectrograph (VIS). The light passing through D2 feeds into the NIR camera ($>$835 nm). The light distribution system (LDS; i.e. M3, D1, D2) and the three channels lie on a single optical plane, enabling their assembly on a common optical bench. Each spectrograph is equipped with a long-slit, a grating, and a CCD detector (see Figure~\ref{fig:optical_dia}).

\begin{figure*}
	\includegraphics[width=0.8\textwidth]{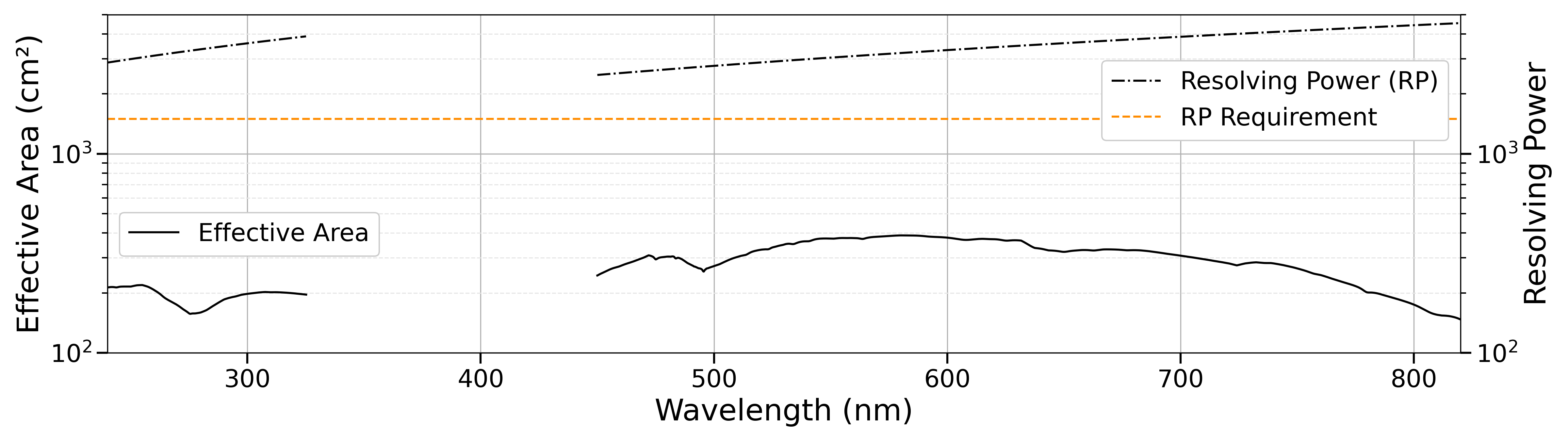}
    \caption{Effective area and resolving power as a function of wavelength for the NUV and VIS spectrographs. The resolving power requirement is shown by a dashed orange line.}
    \label{fig:eff}
\end{figure*}

\noindent{\bf Near-Ultraviolet spectrograph (NUV channel).} The NUV spectrograph is based on that onboard the CUTE SmallSat \citep{Fleming2018,France2023}. The spectrograph provides continuous coverage in the 240--325 nm range at a resolving power $>$2850. The design consisting of a slit, a blazed aberration-correcting holographic concave grating, a cylindrical fold mirror, and a detector. The design minimises reflections to maximise throughput. Wavelength dispersion occurs through a grating that disperses, magnifies, and focuses the light beam. The $25\times25$ mm grating will be made of Zerodur by Horiba. A cylindrical fold mirror after the grating compresses the cross-dispersion profile, improving sensitivity at the shortest wavelengths where the faintest targets are readout-noise limited. However, this fold mirror could be removed with a redesign of the NUV grating, which is currently conceived to be identical to that of CUTE. The removal of the fold mirror would lead to gain of about 10\% in the effective area. The baseline detector of choice is the CCD42-10 from Teledyne e2v ($2048\times515$ pixels, $13.5 \mu\mathrm{m}$ pitch), and the optical design is sized accordingly. Figure~\ref{fig:eff} shows the effective area and resolving power of the NUV spectrograph (with the fold mirror).

\begin{figure*}
	\includegraphics[width=0.8\textwidth]{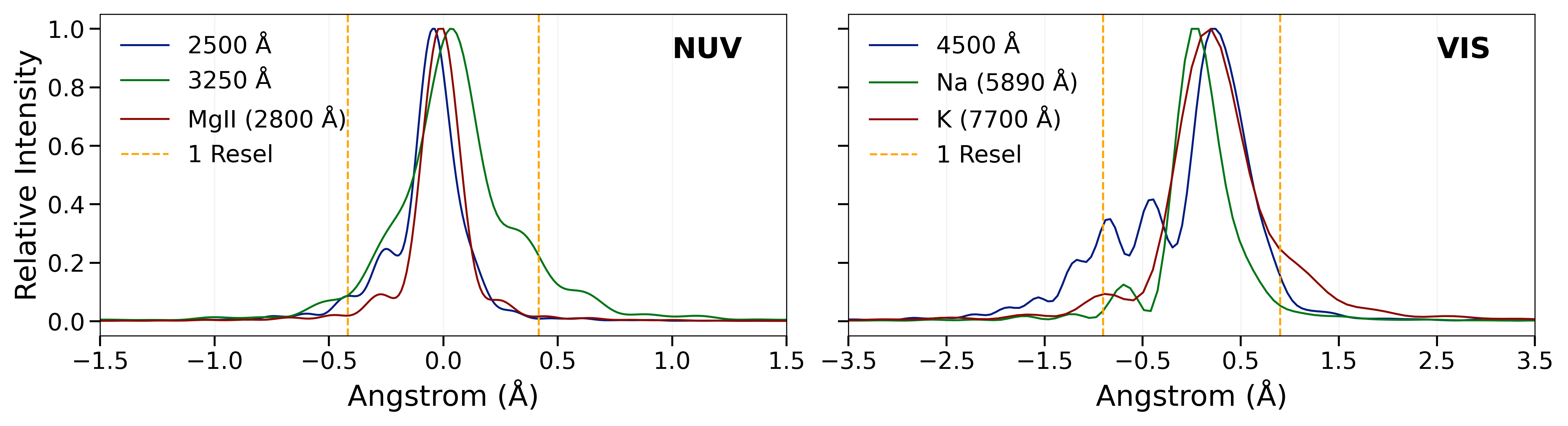}
    \caption{Line spread function at key wavelengths for the two spectrographs. The vertical dashed orange lines indicate the nominal two-pixel detector sampling scale, corresponding to approximately one spectral resolution element.}
    \label{fig:lsf}
\end{figure*}

\noindent{\bf Visible spectrograph (VIS channel).} The design of the VIS spectrograph draws from the NUV design. It comprises a long-slit and a blazed aberration-correcting holographic concave grating. The beam focused by the grating falls directly on the detector. The $25\times25$ mm grating has a curvature radius of 122.76 mm and a groove density of 598.2 gr/mm, and will be manufactured in Zerodur by Horiba. The design is tailored to a detector with a physical size of $55.29\times27.65$ mm and a pixel size of 13.5 µm, corresponding to the CCD42-80 detector from Teledyne e2v with $4096\times2048$ pixels. The spectrograph has a resolving power $>2500$ across the full band. Figures \ref{fig:eff} and \ref{fig:lsf} (right) show the effective area and line spread function (LSF) of the VIS channel.

\noindent{\bf Near-Infrared camera (NIR channel).} The NIR camera provides reference transit light curves covering the water bands to support WALTzER’s core science in concert with the NUV and VIS spectrographs, and enables high-precision correction of the spacecraft pointing. With a typical frame rate faster than 1 Hz, the NIR camera supports real-time feedback for the closed-loop pointing control (e.g. as on CHEOPS, Ariel, ARRAKIHS). The images acquired by the camera are defocused, reducing sensitivity to pixel-to-pixel variations and improving photometric stability. The PSF of a point-source spreads over a circle with a radius of $\sim$5 pixels within the field of view of $640\times512$ pixels, corresponding to an on-sky projection of $12^{’} \times 9.6^{’}$. The NIR camera shall cover the $\sim$900--1600 nm wavelength range and work at low temperatures (less than $-$35$^{\circ}$C) to optimise the amount of signal and reduce the dark current and noise contributions. The NIR camera is baselined after the design of the TANGO-Carbon camera \citep{Jansen25,Cooney25} of the TANGO mission \citep{Cooney2025,Jansen2024} that is part of ESA's SCOUT program \citep{Borsdorff26}. 

To limit straylight, the telescope will use baffles and diaphragms. An external baffle extending behind M2 prevents direct illumination of M1 from out-of-field sources (e.g. Earth) beyond a 20$^{\circ}$ exclusion angle. All baffle surfaces will be ``black'' over the working wavelength range, with total integrated scattering $<$5\% and near-Lambertian behaviour. Less critical areas may use ESA-standard black anodised aluminium, while critical areas may require e.g. Aeroglaze Z306 or Acktar for low outgassing and improved straylight suppression \citep{Marshall+2014}. Similarly, the three channels will include geometric baffling, absorptive coatings, and control of scattering paths to reduce scattering. 


\subsection{Slits}

\begin{figure}
	\includegraphics[width=\columnwidth]{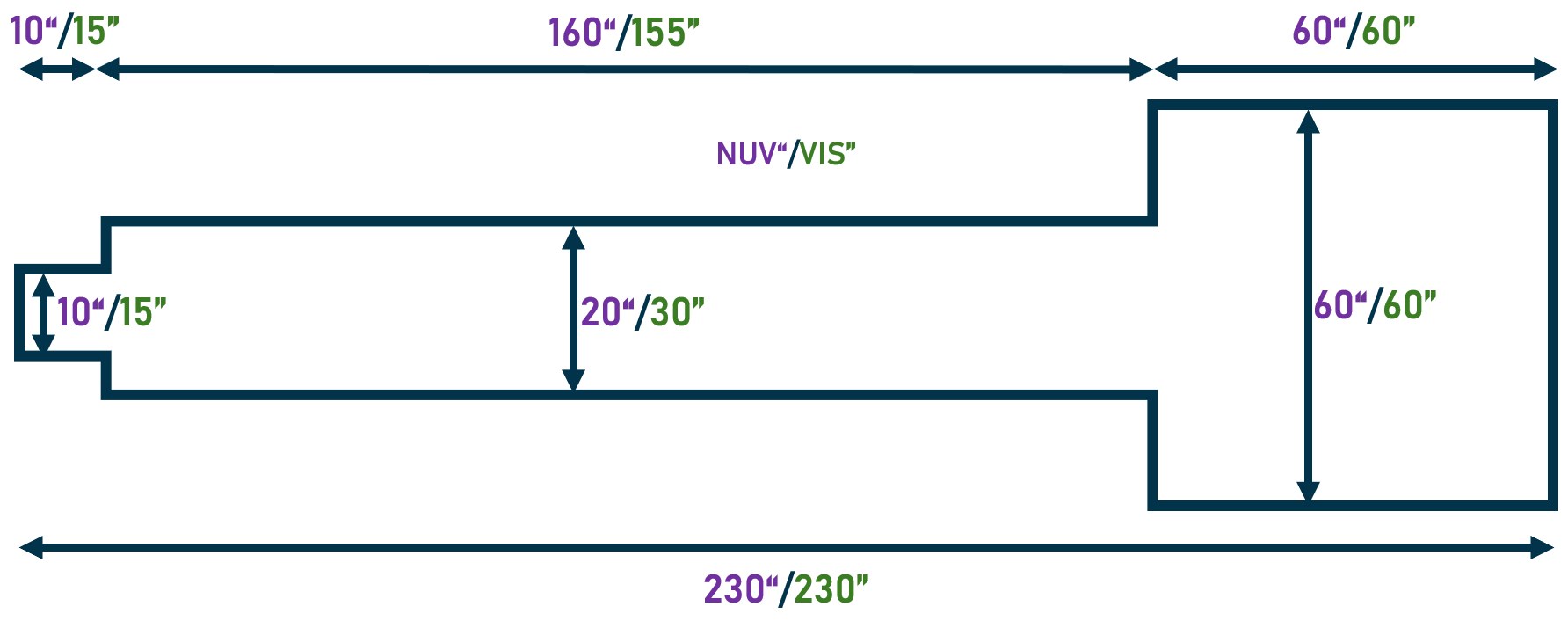}
    \caption{Slit shape and sky-projected dimensions of the NUV (purple) and VIS (green) spectrographs.}
    \label{fig:slit}
\end{figure}

The spectrographs employ identical multi-width long slits that act as field stops. The total slit length of $\sim3.2$ mm (230\arcsec\ on sky) matches the size of the NUV detector in the spatial (i.e. cross-dispersion) direction. Each slit features three distinct widths arranged along the spatial direction (Figure~\ref{fig:slit}). The upper section is the widest and is for observations of extended sources at lower resolving power or of point-like sources that need to be observed without PITL. The lower section is the narrowest and is for targets in crowded fields. The central section is the workhorse slit. Its length allows the target spectrum to be positioned at different locations along the detector, mitigating the impact of hot pixels that may build up over time on the detectors, yet enabling enough room to find reliable background regions. Its width, corresponding to almost 4 times the PSF of a point-like source encircling 100\% of the energy, prevents slit losses accounting for S/C jitter (with PITL) and thermal instabilities. 

\subsection{Detector System}

For both spectrographs, the spectrum is imaged onto e2v CCD42 series detectors. The NUV detector is a back-illuminated, UV-enhanced CCD detector (CCD42-10). The CCD42-10 has an active area of $27.6 \times 6.9$ mm with $2048 \times 515$ pixels, each 13.5 $\mu\mathrm{m}$ square. The CCD will be passively cooled to operate at a temperature of -45$^{\circ}\pm$0.5$^{\circ}$C by a thermal contact with a Cu heat sink and spacecraft radiator system and heaters to maintain the temperature, resulting in an expected dark background rate of $1.2 \times 10^{-2}$ e$^{-}$ pixel$^{-1}$ s$^{-1}$. The expected read noise is around 4 e$^{-}$ pixel$^{-1}$ RMS at the desired readout rate.

The visible channel will employ the larger CCD42-80 detector with a wavelength optimised broadband coating. The detector has an active area of $55.29 \times 27.65$ mm with $4096 \times 2048$ pixels. This detector will operate at the same temperature as the NUV one and will have a similar noise performance. 

For the NIR camera a SNAKE detector from LYNRED will be used for the detection. It consist of InGaAs material with a pixel pitch of 15 $\mu\mathrm{m}$, containing $640 \times 512$ pixels. In a comparative study, this detector has been characterised and subsequently selected for the TANGO mission \citep{Cooney25} at room temperatures. Extrapolating to the planned cryogenic operating temperature of -35$^{\circ}C$, we expect a detector quantum efficiency $>$80\% across most of the targetted wavelength band and a readout noise and dark current of $\leq$300~e$^{-}$ and $\leq$500~e$^{-}$s$^{-1}$, respectively. At these temperatures, the cut-off wavelength of the InGaAs material will be longwards of 1600 nm.  

\subsection{Mechanical and thermal}

The WALTzER mechanical and thermal design builds on the flight-proven CHEOPS architecture, with improvements to enhance thermo-elastic stability. A CFRP/Invar36 optical bench supports the telescope, spectrographs, and NIR camera, with all optical elements mounted on the same plane to minimise alignment sensitivity. Lightweight mirrors, low-CTE structural materials, flexible interfaces, and three-point shimmed mounts ensure high structural stability, while reducing mass and sensitivity to thermal distortions. Baffled optical paths suppress stray light and protect against contamination. The optical bench is actively stabilised near room temperature to facilitate ground calibration, while the detectors are thermally isolated from the structure and cooled via graphite thermal straps, heat pipes, and dedicated radiators. The CCDs operate at $-45^\circ$C and the NIR detector below $-35^\circ$C, with all detectors requiring a temperature stability of $\pm0.1$ K. This stability is achieved through operational heaters, precision temperature sensors, and thermal capacitors, an approach demonstrated by the CHEOPS mission \citep{Deline2020}, which maintains detector temperatures to within a few mK \citep{fortier2024}.

\subsection{Electronics}

The electronic box (EBOX) inside the platform controls the instrument. The EBOX is a slotted-frame design containing 8 electronic boards: 2 Power Supply Units (PSU), 2 Data Processing Units (DPU), 2 Thermal Control Units (TCU), in each case a nominal (-N) and a redundant (-R) unit. There are also 2 Back-End Electronics (BEE) for the NUV and VIS channels. The EBOX is a cold-chain design, where PSU-N interfaces with DPU-N and TCU-N, while PSU-R interfaces with DPU-R and TCU-R. The BEEs are not redundant because they are an extension of each detector chain, and will accommodate two communication connectors (DPU-N and DPU-R). The OR-ing of the PSU-N and PSU-R power is performed directly on the PSU output, with diodes. Finally, the DPUs and PSUs interface externally to the NIR Front end electronics (FEE), located near the NIR sensor that also supports two communication and one power connector. The front-end electronics will be based on the CCD readout chain \citep{Giono2018} used onboard Swedish MATS satellite \citep{Gumbel2020}."

\subsection{Instrument performance}

WALTZER achieves an average resolving power greater than 2850/2500 across the operational bandpass for the NUV/VIS channel, respectively. The spectral resolving power is determined by the effective spectral resolution element, which is nominally sampled by two CCD pixels (27~$\mu\mathrm{m}$). The optical design maintains an RMS spot width in the dispersion direction below this scale at each wavelength, such that the instrumental line-spread function is adequately sampled. Figure~\ref{fig:lsf} shows spectral LSFs at three representative wavelengths across the detector, with the width of one resolution element indicated by vertical bars. The LSF describes the response of a spectrograph to an infinitely narrow monochromatic spectral line and the instrument operates close to the Nyquist sampling criterion, preserving the designed resolving power.

\subsubsection{Effective Area}

The effective area (Figure~\ref{fig:eff}) is the product of the clear collecting area of the telescope ($A_T$) and the efficiencies of the individual optical and detector components, that is
\begin{align}
A_{\mathrm{eff}}^{\mathrm{NUV}}(\lambda) &=
A_T \times Rf(\lambda)^5 \times Rf_{D1}(\lambda)
\times G_{\mathrm{eff}}^{\mathrm{NUV}}(\lambda)
\times D_{\mathrm{QE}}^{\mathrm{NUV}}(\lambda), \\[6pt]
A_{\mathrm{eff}}^{\mathrm{VIS}}(\lambda) &=
A_T \times Rf(\lambda)^4 \times T_{D1}(\lambda)
\times Rf_{D2}(\lambda)
\times G_{\mathrm{eff}}^{\mathrm{VIS}}(\lambda)
\times D_{\mathrm{QE}}^{\mathrm{VIS}}(\lambda), \\[6pt]
A_{\mathrm{eff}}^{\mathrm{NIR}}(\lambda) &=
A_T \times Rf(\lambda)^3 \times T_{D1}(\lambda)
\times T_{D2}(\lambda)
\times D_{\mathrm{QE}}^{\mathrm{NIR}}(\lambda).
\end{align}
Here, $Rf(\lambda)$ is the reflectivity of the MgF$_2$-protected Al-coated optics; $Rf_{D1}(\lambda)$, $Rf_{D2}(\lambda)$, $T_{D1}(\lambda)$, and $T_{D2}(\lambda)$ are the manufacturers' reflectivity and transmissivity of the dichroics; and $G_{\mathrm{eff}}(\lambda)$ and $D_{\mathrm{QE}}(\lambda)$ are the manufacturers' grating efficiency and the detectors' quantum efficiency, respectively.

\begin{figure*}
	\includegraphics[width=\textwidth]{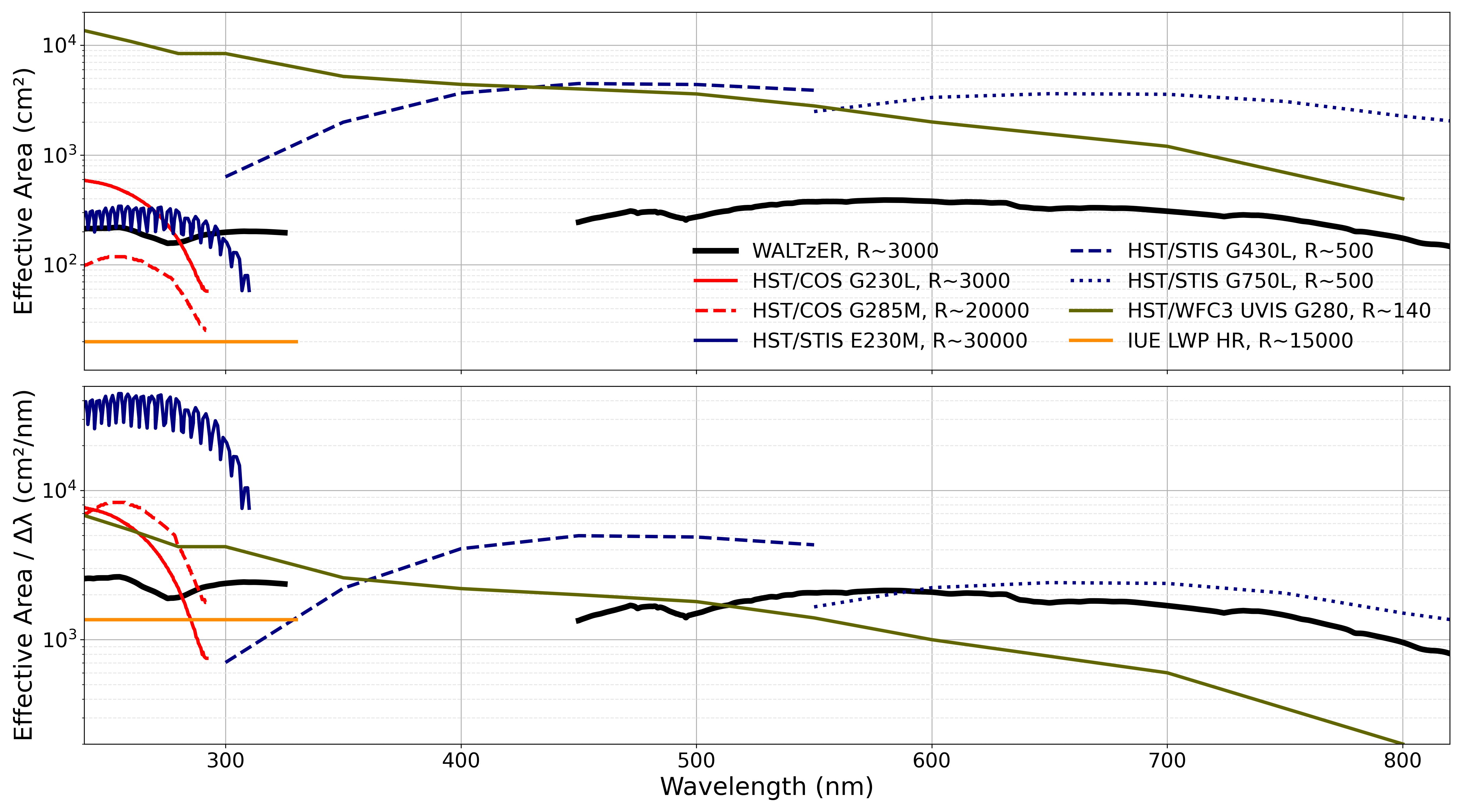}
    \caption{Effective Area (EA; top) and EA per resolution element (bottom) curves of WALTzER (\diameter 35 cm) and relevant modes of HST/STIS, HST/COS, HST/WFC3 (\diameter 2.4 m), and IUE (\diameter 45 cm; only peak EA is shown). Note: HST/COS G230L and G285M cannot cover the entire WALTzER NUV band at once, but each requires using 3 separate settings.}
    \label{fig:eff_comp}
\end{figure*}

Figure~\ref{fig:eff_comp} compares the effective area and the effective area per resolution element of WALTzER with that of the workhorse space-based instruments used to carry out NUV and visible exoplanet transmission spectroscopy observations \citep[e.g.][]{Fossati+2010,Sing+2011,Sing2016,Sing2019,Cubillos2020,Cubillos2023,Wakeford2020,Lothringer2022,Baldwin2026}. Despite the much larger HST mirror (\diameter 2.4 m vs \diameter 35 cm), WALTzER’s effective area is competitive with that of various HST/COS, STIS, and WFC3 instrument modes. WALTzER's effective area in the NUV is significantly greater than that of IUE.

\subsubsection{Noise budget}

To evaluate the achievable performance in the three channels, we developed a detailed noise budget in which the noise sources are grouped into three main categories: random, systematic, and environmental noise. The calculation is performed for representative wavelengths of 3000~\AA, 5500~\AA, and 12,500~\AA\ for the NUV, VIS, and NIR channels, respectively. The corresponding source count rates are obtained from the ETC described earlier, and the noise contributions are expressed in parts per million (ppm) relative to the detected stellar signal.

\textit{Random noise} includes photon (shot) noise from the target, zodiacal and Earth-shine background, detector read noise, dark-current noise, and detector quantisation noise. These are added in quadrature. The photon noise is calculated directly from the number of detected source electrons in each exposure
\begin{equation}
\sigma_{\rm ph} = \frac{10^6}{\sqrt{N_{\rm src}}},
\end{equation}
where $N_{\rm src}$ is the number of detected source electrons. For a six-hour observation, the resulting uncertainty decreases as the square root of the number of exposures. We assume exposure times of 300~s in the NUV and VIS channels and 60~s in the NIR channel, corresponding to 72, 72, and 360 exposures, respectively, over six hours.

The detector-related random noise terms are calculated using the number of pixels contributing to the spectral extraction or photometric aperture. These values are obtained based on the optical design. The detector noise information is based on the data sheets provided by the manufacture. Both the source and background regions are included in the propagation of the read-out and dark-current uncertainties. The quantisation noise is calculated assuming 16-bit digitisation and a full-well capacity of $10^5$ electrons. The resulting instrumental noise is obtained by adding the read-out, dark-current, and quantisation terms in quadrature.

The zodiacal-light and Earth-shine contributions are treated as background signals and are propagated using the number of background pixels and the pixels in the source extraction region. The environmental noise budget additionally includes straylight and pointing jitter. The straylight level is limited to 0.1~photons~pixel$^{-1}$~s$^{-1}$, with appropriate channel-dependent coupling factors introduced. The resulting straylight uncertainty is propagated to the extracted source signal. Pointing jitter is represented by a fractional photometric perturbation of 0.1\% in each channel and is scaled according to the number of pixels contributing to the measurement. The jitter and straylight contributions are then combined in quadrature to obtain the total environmental noise.

\textit{Systematic noise} represents residual instrument-related variations associated with detector and instrument stability, including temperature-dependent changes in dark current, quantum efficiency, and detector gain. For the detector noise terms, the contribution from the background estimate is included in the propagation of the systematic uncertainty. The adopted values represent the residual systematic performance expected after calibration and pre-flight characterisation.

For each magnitude, the total single-exposure noise is calculated by adding the independent contributions in quadrature,
\begin{equation}
\sigma_{\rm tot} =
\left(
\sigma_{\rm sys}^2 +
\sigma_{\rm env}^2 +
\sigma_{\rm ZL}^2 +
\sigma_{\rm instr}^2 +
\sigma_{\rm ph}^2
\right)^{1/2}.
\end{equation}
For the six-hour observing scenario, the random and environmental contributions are reduced according to the square root of the number of exposures. In the present implementation, the adopted systematic term is also propagated using this same $1/\sqrt{N_{\rm exp}}$ scaling. Fully non-averaging systematic offsets are not included in this six-hour reduction and would instead constitute a systematic noise floor.

\begin{figure*}
	\includegraphics[width=\textwidth]{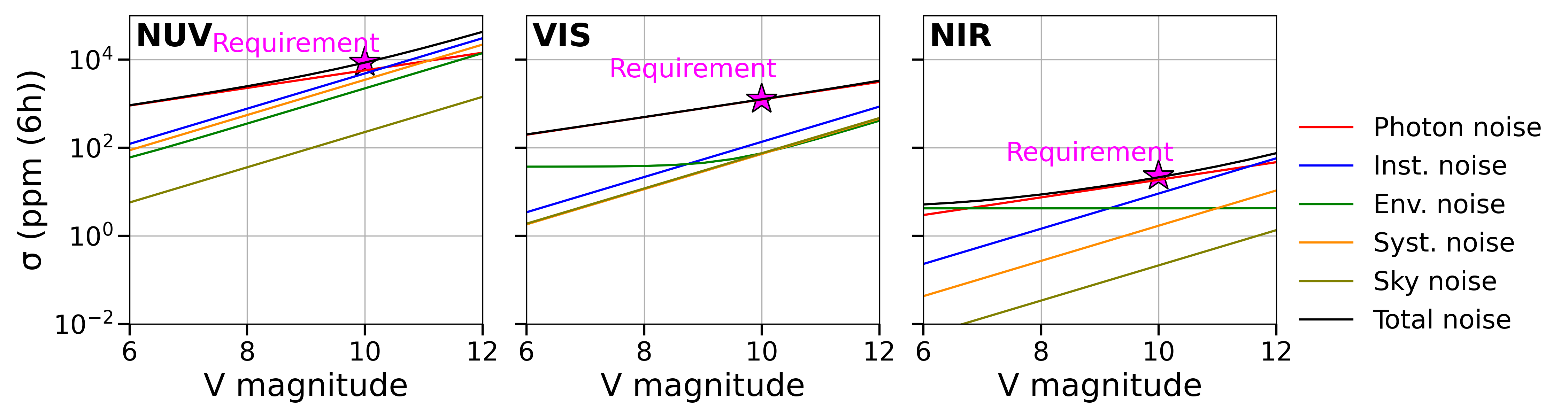}
    \caption{WALTzER’s noise budget for the three channels. The requirements for a 6000 K Sun-like star at V=10 mag (Section~\ref{sec:requirements}) are shown in magenta. The NUV and VIS noises are calculated per pixel, whereas the NIR noises corresponds to broadband photometry.}
    \label{fig:waltzer_noise_plot}
\end{figure*}

Figure~\ref{fig:waltzer_noise_plot} shows the noise behaviour as a function of target magnitude for a fixed 6-hour observation. The instrument is photon noise dominated for all targets with V$\le10$ mag (for faint targets other noise sources dominate), except for bright objects in the NIR channel as a result of flat field errors, assumed to be of 0.1\%.

\subsubsection{Pointing performance}

The expected fine-pointing performance was assessed through end-to-end simulations of the NIR imager conducted using the WALTzER data simulator developed based on \citet{Sreejith2019,Sreejith2023}. The simulator generates realistic detector images by modelling the telescope optics, point spread function, detector characteristics, photon and detector noise, and the observing conditions expected during the mission. A large set of synthetic observations was produced, covering the range of stellar properties relevant for the mission, including different spectral types, apparent magnitudes, and stellar parameters representative of the WALTzER target sample.

The simulated images were processed using a centroid algorithm closely based on the one successfully employed by the CHEOPS mission \citep{futyan2020}. For each simulated observation, the estimated stellar centroid was compared with the true source position, and the centroid error was recorded. A Monte Carlo analysis was then performed to quantify the expected pointing accuracy as a function of the detected signal level.

\begin{figure}
	\includegraphics[width=\columnwidth]{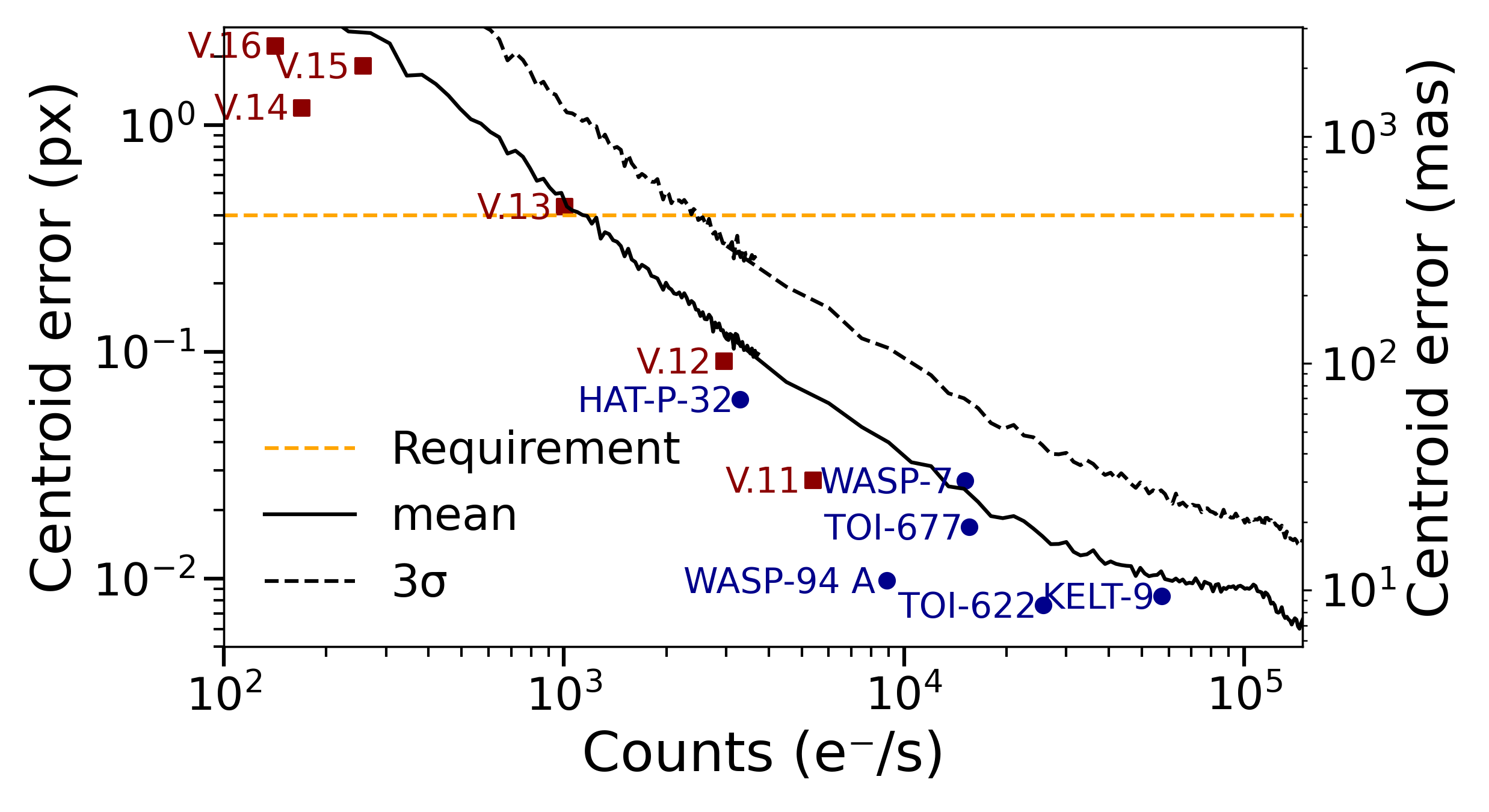}
    \caption{Simulation of the centroid estimation error for point-like sources. The black solid and dashed lines show the mean error and 99.7\% envelope based on a Monte Carlo simulation. Labelled points mark the errors obtained for simulated observations of WALTzER targets (blue) and Sun-like sources at different V-band magnitudes (red). For 1 s integrations, the guiding requirement of 0.44\arcsec\ is met up to V=13 mag (V=12.5 mag using the 99.7\% criterion), while for fainter targets longer integrations (up to 10 s) are needed to yield the required signal of $\sim$2500 e$^-$.}
    \label{fig:pointing}
\end{figure}

The results are presented in Figure~\ref{fig:pointing}. The black solid line shows the mean centroid error, while the dashed lines indicate the 99.7\% confidence envelope obtained from the Monte Carlo simulations. Labelled points correspond to representative WALTzER science targets (blue) and simulated Sun-like stars with different $V$ magnitudes (red; mimicking comets of different magnitudes), illustrating the expected performance across the mission's target brightness range.

The simulations demonstrate that, for 1 s integrations, the centroid requirement of 0.44\arcsec\ is satisfied for targets brighter than $V=13$ mag, or up to $V=12.5$ mag when adopting the more conservative 99.7\% confidence criterion. For fainter targets, the centroid accuracy becomes photon-noise limited and can be recovered by increasing the integration time up to $\sim$10 s in order to reach the required signal level of approximately 2500 detected photoelectrons.

\section{Science operations}\label{sec:science_operations}

WALTzER targets are distributed across the entire sky, requiring a mission architecture that maximises sky accessibility, while minimising systematic noise in time-series observations building on lessons from previous space-based transit observations \citep[e.g.][]{Gibson2011,Sing2019,Sreejith2023b,Fossati2025}. Therefore, the instrument combines three channels within a single payload, avoids slit losses, performs simultaneous spectroscopy and photometry, and adopts a thermally stable spacecraft attitude that shields the radiators from Sun and Earth illumination. 

The baseline mission uses a dawn-dusk low-Earth orbit (6 am - 6 pm) at an altitude of $\sim$700 km. No orbit transfer is required after launch, and the mission is compatible with both full and shared Vega-C launches, with a full Vega-C launch adopted as the baseline. The spacecraft is based on a CHEOPS-like platform equipped with GPS for precise orbit determination and target visibility prediction, together with X-band communications. 

\begin{figure}
	\includegraphics[width=\columnwidth]{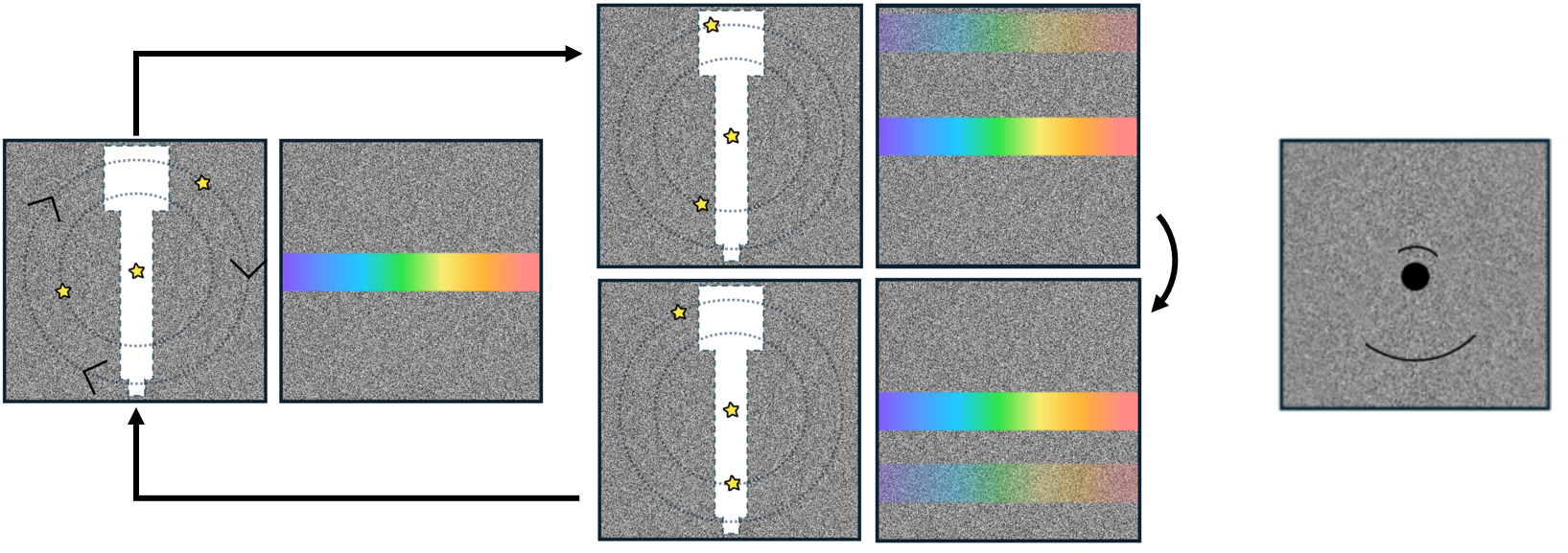}
    \caption{Illustration of the timeline of the motion of the field of view around a target as it would be shown by the NIR camera (right) after windowing and by one of the spectrographs (left and middle). As background objects briefly pass through the slit, they may leave a faint and smeared spectrum on the detector.}
    \label{fig:rotating_field}
\end{figure}

The spacecraft will be Nadir-locked to ensure thermal stability by preventing direct illumination of radiators by the Earth. A Nadir-locked configuration leads to a rotating field of view, meaning that celestial objects will rotate around the target with a period equal to the S/C orbital period ($\sim$100 min). In the NUV and VIS channels, the spectrum of the target is locked in position on the detectors, while spectra of bright enough objects surrounding the target may appear on the detector when they fall inside the slit (Figure~\ref{fig:rotating_field}). Among the core science targets, just two (HAT-P-1 and HD\,189733) lie in fields crowded enough to possibly pose a problem. For maximum sky coverage, we require a Sun exclusion angle of $110^{\circ}$, which can be achieved with proper baffling of the radiators. To minimise straylight, we require an Earth-limb exclusion angle between $20^{\circ}$ and $30^{\circ}$ (average of $25^{\circ}$) and a Moon exclusion angle of $5^{\circ}$. 

\begin{figure*}
	\includegraphics[width=\textwidth]{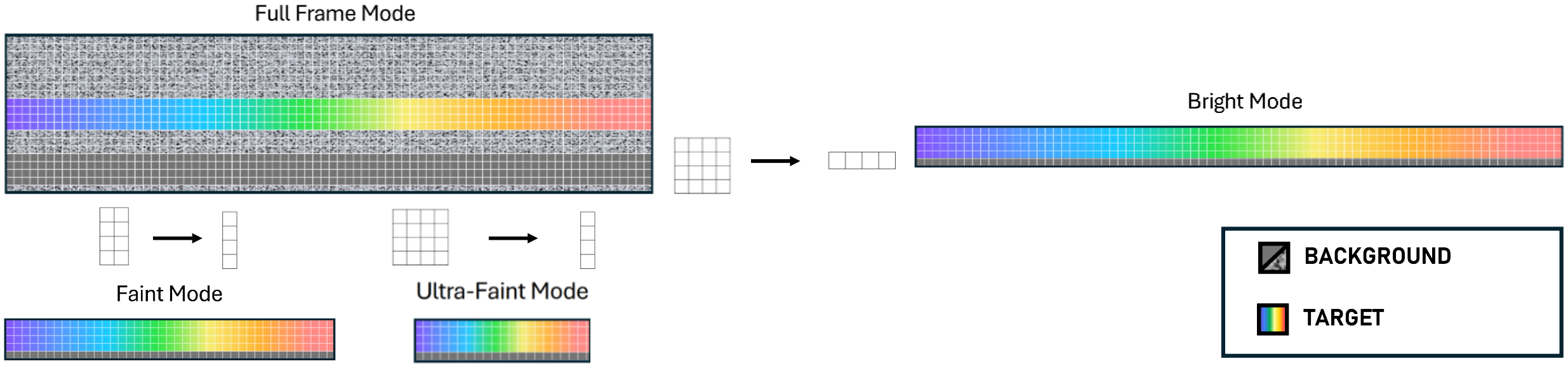}
    \caption{Illustration of the different observing modes available for the spectroscopic channels.}
    \label{fig:modes}
\end{figure*}

The three channels operate simultaneously, although each can be assigned an independent exposure time. The NUV and VIS channels support four readout modes (Figure~\ref{fig:modes}): full-frame mode (FFM), bright mode (BRM), faint mode (FNM), and ultra-faint mode (UFM). FFM reads and downlinks the entire CCD and is primarily used for commissioning, monitoring and calibration (M\&C), or observations of crowded fields. BRM is the nominal science mode, reading the target spectrum pixel-by-pixel, while applying on-chip binning to the background region in the cross-dispersion direction to reduce readout time and noise. FNM and UFM further reduce readout noise through on-chip binning in the dispersion direction by factors of two and four, respectively. FNM preserves the nominal resolving power ($R\sim3000$), whereas UFM reduces it to $R\sim1500$. These modes are suitable for faint targets.

The NIR channel supports two modes: FFM and window mode (WDM). FFM reads the full detector and is reserved for commissioning and M\&C activities. In WDM, a $\sim210\times210$ pixel window centred on the target is read out with short (1--10 s) integrations that also provide the fine-guidance signal. The onboard software co-adds these exposures into 1 min science frames before downlink. WDM is the standard observing mode available to users.

A typical observation consists of slewing to the target, using the NIR channel for target acquisition and fine pointing, executing the science exposure sequence, and then slewing to the next target.

Core science observations are carried out during exoplanet transits. Each observation covers the transit (typically 1--5 hours) plus pre- and post-transit phases to set the stellar flux baseline. To this end each observation should be at least as long as three times the transit duration and always ensuring to cover the pre- and post-transit phases for the duration of a minimum of three satellite orbits. Typical observations of the secondary surveys are single pointing, which total duration depends on the target’s magnitude, with the longer exposures split in repeated observations. WALTzER observations mix time-constrained and non-time-constrained observations. Therefore, the consortium will develop a Mission Planning System (MPS) optimising the observing sequence to maximise the use of telescope time. The MPS will use heritage of projects already in operation \citep[e.g.][]{Garcia-Piquer2017} and in development \citep[e.g.][]{Morales2022}. It will receive observation requests as input and search for the best combination of non-overlapping transit observations fulfilling the target observability constraints (given the payload capabilities), while maximising the number of targets for which all observations are completed. Time-constrained observations (e.g. transits) will drive the mission plan generation, while idle time slots between transits will be used to schedule non-time-constrained observations.

\section{Summary and conclusions}\label{sec:conclusions}

WALTzER has been conceived as a dedicated space mission for high-precision, simultaneous ultraviolet, visible, and near-infrared spectro-photometry of transiting exoplanets. Its design addresses several of the principal limitations encountered so far by space-based atmospheric characterisation observations: WALTzER combines broad spectral coverage, simultaneous multi-band observations, and high photometric stability within a single instrument.

The optical design provides continuous wavelength coverage across three channels while maximising throughput through carefully selected optical coatings, dichroics, gratings, and detectors. A flexible detector readout architecture allows the observing strategy to be optimised for targets spanning a wide range of apparent magnitudes without compromising spectral performance. End-to-end performance simulations, together with the dedicated ETC, demonstrate that the instrument is capable of delivering the sensitivity and precision required to carry out its core science program.

The mission concept further minimises systematic effects through a thermally stable dawn-dusk low-Earth orbit, a spacecraft platform inherited from CHEOPS, and an integrated fine-guidance system based on the NIR detector. Realistic image simulations show that the pointing requirements can be met over essentially the entire target sample. The combination of simultaneous multi-band observations, optimised observing modes, efficient scheduling, and a stable spacecraft platform make WALTzER a powerful facility for precision exoplanet atmospheric characterisation and time-domain stellar and Galactic astrophysics.

In the 2030s, JWST and the ELT will dominate the astrophysical landscape across disciplines, alongside numerous other facilities that will continue advancing the field. However, no existing or planned facility will provide NUV-to-NIR simultaneous access. WALTzER will therefore provide a unique capability, delivering datasets that bridge large survey missions and detailed characterisation facilities, while creating a cost-effective and lasting scientific legacy.

\section*{Acknowledgements}

HC acknowledges support from the Swiss National Science Foundation (SNSF) under the Postdoc Mobility grant P500-2\_239025. J.C.M. and I.R. acknowledge financial support from Spanish grants PID2024-158486OB-C31 funded by MCIU/AEI/10.13039/501100011033 and by “ERDF A way of making Europe”, from the programme Unidad de Excelencia María de Maeztu CEX2020-001058-M, from the MaX-CSIC Excellence Award MaX4 SOMMA-ICE, and from the Generalitat de Catalunya/CERCA programme. JZ acknowledges the support of the PRODEX agreement no. 400012791.KK acknowledges the support by the European Union (ERC, EASE, 101123041). Views and opinions expressed are however those of the author(s) only and do not necessarily reflect those of the European Union or the European Research Council Executive Agency. Neither the European Union nor the granting authority can be held responsible for them. JMC work was (partially) supported by the Spanish MICIN/AEI/10.13039/501100011033 and by "ERDF A way of making Europe" by the “European Union” through grant PID2021-122842OB-C21 and PID2024-157964OB-C21, and the Institute of Cosmos Sciences University of Barcelona (ICCUB, Unidad de Excelencia ’Mar\'{\i}a de Maeztu’) through grant CEX2024-001451-M and the project 2021-SGR-00679 GRC de l'Agència de Gestió d'Ajuts Universitaris i de Recerca (Generalitat de Catalunya). TGW acknowledges support from the UKSA and University of Warwick. C.D. acknowledges financial support from the grant RYC2023-044903-I funded by both MCIU/AEI/10.13039/501100011033 and the ESF+. V.V.G. Is a F.R.S.-FNRS Senior Research Associate. MB is supported by an Ambizione grant (PZ00-2\_240629) from the Swiss National Science Foundation (SNSF). ASM acknowledges support from the Spanish Ministry of Science and Innovation (MICINN) under the 2024 Ramón y Cajal program MICINN
RYC2024-050707-I funded by MICIU/AEI/10.13039/501100011033 and the FSE+,
and the European Union under the CELESTE project (EU WIDERA Teaming, GA
101136736.PK and JZ acknowledge support from
ESA PEA4000127913. This work has been carried out within the framework of the National Centre of Competence in Research PlanetS supported by the Swiss National Science Foundation under grant 51NF40\_205606. A.De. acknowledges the financial support of the SNSF. Part of the simulated results were obtained using the Austrian Scientific Computing (ASC) infrastructure.


\section*{Data Availability}
This work makes use of open-access data.  WALTzER ETC is available at \href{https://github.com/pcubillos/waltzer_etc}{https://github.com/pcubillos/waltzer\_etc}.


\bibliographystyle{rasti}
\bibliography{waltzer} 




\appendix


\bsp	
\label{lastpage}
\end{document}